\documentclass[
 aps,pra,showpacs,superscriptaddress,twocolumn,amsmath,amssymb,floatfix,
 reprint,%
]{revtex4-1}
\usepackage{xcolor}
\usepackage[
	pdffitwindow=true,
	colorlinks=true,
	frenchlinks=false,
        linkcolor=blue,
	anchorcolor=blue,
        citecolor=blue,
        filecolor=blue,
        urlcolor=blue,
        bookmarks=true,
        bookmarksopen=true,
	bookmarksnumbered=true, 
        bookmarksopenlevel=1,
        plainpages=false,
        pdfpagelabels=true,
	breaklinks
]{hyperref}
\usepackage[per-mode=symbol,separate-uncertainty]{siunitx}
\usepackage{graphicx}
\usepackage{dcolumn}
\usepackage{bm}
\usepackage[english]{babel}
\usepackage{cases}

\begin{document}

\preprint{AIP/123-QED}

\title[]{Second-order decoherence mechanisms of a transmon qubit\\probed with thermal microwave states}

\author{J.~Goetz}
\email[]{jan.goetz@wmi.badw.de}
\affiliation{Walther-Mei{\ss}ner-Institut, Bayerische Akademie der Wissenschaften, 85748 Garching, Germany }
\affiliation{Physik-Department, Technische Universit\"{a}t M\"{u}nchen, 85748 Garching, Germany}
\author{F.~Deppe}
\affiliation{Walther-Mei{\ss}ner-Institut, Bayerische Akademie der Wissenschaften, 85748 Garching, Germany }
\affiliation{Physik-Department, Technische Universit\"{a}t M\"{u}nchen, 85748 Garching, Germany}
\affiliation{Nanosystems Initiative Munich (NIM), Schellingstra{\ss}e 4, 80799 M\"{u}nchen, Germany}
\author{P.~Eder}
\affiliation{Walther-Mei{\ss}ner-Institut, Bayerische Akademie der Wissenschaften, 85748 Garching, Germany }
\affiliation{Physik-Department, Technische Universit\"{a}t M\"{u}nchen, 85748 Garching, Germany}
\affiliation{Nanosystems Initiative Munich (NIM), Schellingstra{\ss}e 4, 80799 M\"{u}nchen, Germany}
\author{M.~Fischer}
\affiliation{Walther-Mei{\ss}ner-Institut, Bayerische Akademie der Wissenschaften, 85748 Garching, Germany }
\affiliation{Physik-Department, Technische Universit\"{a}t M\"{u}nchen, 85748 Garching, Germany}
\affiliation{Nanosystems Initiative Munich (NIM), Schellingstra{\ss}e 4, 80799 M\"{u}nchen, Germany}
\author{M.~M{\"u}ting}
\affiliation{Walther-Mei{\ss}ner-Institut, Bayerische Akademie der Wissenschaften, 85748 Garching, Germany }
\affiliation{Physik-Department, Technische Universit\"{a}t M\"{u}nchen, 85748 Garching, Germany}
\author{J.~{Puertas Mart{\'i}nez}}
\affiliation{Universite Grenoble Alpes, Institut NEEL, F-38000 Grenoble, France}
\affiliation{CNRS, Institut NEEL, F-38000 Grenoble, France}
\author{S.~Pogorzalek}
\affiliation{Walther-Mei{\ss}ner-Institut, Bayerische Akademie der Wissenschaften, 85748 Garching, Germany }
\affiliation{Physik-Department, Technische Universit\"{a}t M\"{u}nchen, 85748 Garching, Germany}
\author{F.~Wulschner}
\affiliation{Walther-Mei{\ss}ner-Institut, Bayerische Akademie der Wissenschaften, 85748 Garching, Germany }
\affiliation{Physik-Department, Technische Universit\"{a}t M\"{u}nchen, 85748 Garching, Germany}
\author{E.~Xie}
\affiliation{Walther-Mei{\ss}ner-Institut, Bayerische Akademie der Wissenschaften, 85748 Garching, Germany }
\affiliation{Physik-Department, Technische Universit\"{a}t M\"{u}nchen, 85748 Garching, Germany}
\affiliation{Nanosystems Initiative Munich (NIM), Schellingstra{\ss}e 4, 80799 M\"{u}nchen, Germany}
\author{K.~G.~Fedorov}
\affiliation{Walther-Mei{\ss}ner-Institut, Bayerische Akademie der Wissenschaften, 85748 Garching, Germany }
\affiliation{Physik-Department, Technische Universit\"{a}t M\"{u}nchen, 85748 Garching, Germany}
\author{A.~Marx}
\affiliation{Walther-Mei{\ss}ner-Institut, Bayerische Akademie der Wissenschaften, 85748 Garching, Germany }
\author{R.~Gross}
\email[]{rudolf.gross@wmi.badw.de}
\affiliation{Walther-Mei{\ss}ner-Institut, Bayerische Akademie der Wissenschaften, 85748 Garching, Germany }
\affiliation{Physik-Department, Technische Universit\"{a}t M\"{u}nchen, 85748 Garching, Germany}
\affiliation{Nanosystems Initiative Munich (NIM), Schellingstra{\ss}e 4, 80799 M\"{u}nchen, Germany}

\date{prel.~version~\today}

\begin{abstract}
Thermal microwave states are omnipresent noise sources in superconducting quantum circuits covering all relevant frequency regimes. We use them as a probe to identify three second-order decoherence mechanisms of a superconducting transmon qubit. First, we quantify the efficiency of a resonator filter in the dispersive Jaynes-Cummings regime and find evidence for parasitic loss channels. Second, we probe second-order noise in the low-frequency regime and demonstrate the expected $T^{3}$ temperature dependence of the qubit dephasing rate. Finally, we show that qubit parameter fluctuations due to two-level states are enhanced under the influence of thermal microwave states. In particular, we experimentally confirm the $T^{2}$-dependence of the fluctuation spectrum expected for noninteracting two-level states.
\end{abstract}

\maketitle

\section{Introduction}
Solid-state based quantum circuits are attractive systems for quantum information and quantum eletrodynamics (QED) due to their design flexibility and the possibility to engineer and tune interactions. This is particularly true for superconducting quantum circuits which are widely used for quantum computing~\cite{Lucero_2012} and quantum simulation~\cite{Houck_2012}, or for the generation of quantum entanglement~\cite{Fedorov_2016}. One advantage of superconducting circuits is that they provide strong~\cite{Wallraff_2004,Zollitsch_2015} or even ultrastrong~\cite{Niemczyk_2010,Forn-Diaz_2010,Baust_2016} and well controllable~\cite{Hime_2006,Niskanen_2007,Bialczak_2011,Baust_2015,Wulschner_2016} interaction. However, while strong interaction enables simple and fast manipulation of quantum circuits, it goes along with strong coupling to environmental fluctuations (noise), thereby limiting the coherence properties. For superconducting quantum circuits, the impact of environmental noise has been widely studied both in theory and experiment. In particular, noise sources that couple coherently to qubits~\cite{Simmonds_2004,Shalibo_2010,Lisenfeld_2010}, as well as Markovian~\cite{Reed_2010,Kim_2011,Zaretskey_2013,Bronn_2015,Haeberlein_2015}, or non-Markovian $(1/f)$ noise sources~\cite{Astafiev_2006,Yoshihara_2006,Koch_2007a,Schreier_2008,Burkard_2009,Bylander_2011} have been analyzed. To optimize the coherence properties, several strategies to decouple a qubit from the environmental noise have been developed. In the first place, the most convenient way to suppress noise over a broad frequency range is to place the qubit inside a superconducting resonator~\cite{Houck_2008}. This concept is efficient when the qubit transition frequency is far detuned from the resonator frequency by an amount $\delta$ much larger than their coupling strength $g$. Nevertheless, even in this case, noise still couples to the qubit in second-order with strength $g^{2}/\delta$. In the second place, fluctuations that modify the qubit transition frequency $\omega_{\mathrm{q}}$ can be noticeably suppressed by tuning the qubit to an operation point where the derivative of $\omega_{\mathrm{q}}$ with respect to the fluctuating quantity vanishes~\cite{Vion_2002,Yoshihara_2006,Koch_2007,Schreier_2008,Bylander_2011}. Again, even at such a sweet spot, second-order coupling of environmental fluctuations can be a source for decoherence~\cite{Shnirman_2002,Makhlin_2003,Makhlin_2004}. In addition to these decoherence processes, intrinsic qubit parameters such as its relaxation rate can be fluctuating in time~\cite{Paik_2011,Yan_2012,Muller_2015,Yan_2015,Gustavsson_2016}. One prominent source for these fluctuations are two-level states (TLSs) mediating low-frequency noise to the qubit.

The three second-order decoherence mechanisms mentioned above can be reliably studied with propagating thermal fields because their power spectral density $\mathcal{S}(\omega)$ can be adjusted with a high accuracy by controlling the temperature of a black-body radiator~\cite{Mariantoni_2010,Menzel_2010,Menzel_2012}. Furthermore, $\mathcal{S}(\omega)$ is white for low frequencies and sufficiently smooth at the qubit transition frequency, which allows for a quantitative analysis of second-order decoherence mechanisms. Besides the fact that thermal fields are an accurate control knob to study second-order effects of noise, their omnipresence in superconducting circuits~\cite{Bertet_2005a,Corcoles_2011,Sage_2011,Sears_2012,Rigetti_2012,Pop_2014,Goetz_2016} naturally results in a strong demand to investigate their second-order influence on the coherence properties of superconducting quantum circuits.

\begin{figure*}[t]
\includegraphics{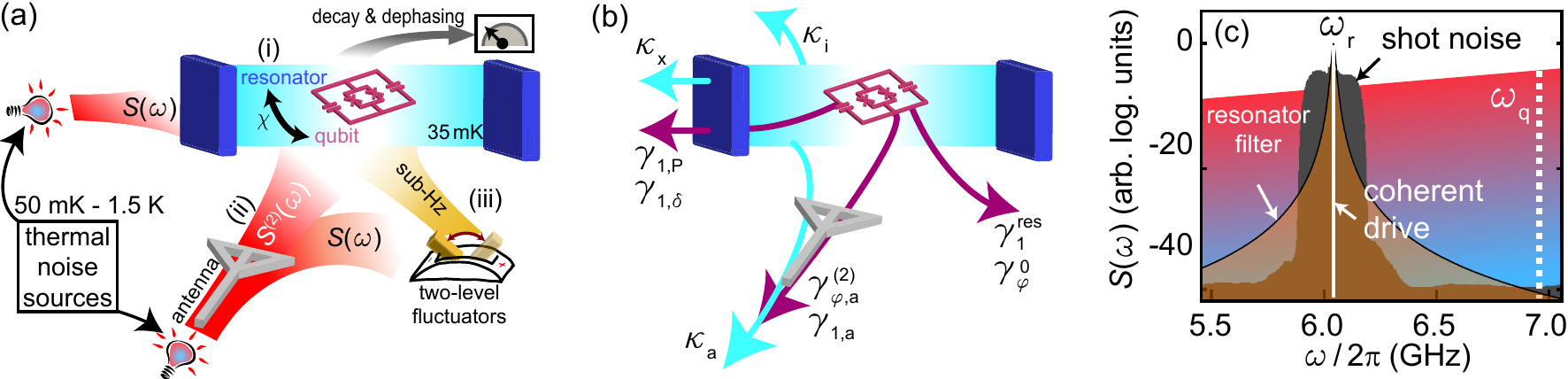}
\caption{\label{fig:scheme}(a) Sketch of the experimental idea. We characterize the second-order coupling between a thermal noise source (black-body radiator of variable temperature) and a superconducting qubit by measuring the qubit decay and dephasing rate. The second-order coupling to the thermal noise field becomes relevant in the following situations: (i) When the qubit is dispersively coupled to a resonator acting as a filter for the noise field with spectral density $\mathcal{S}(\omega)$. (ii) When the qubit is directly irradiated by a noise field using a near-field antenna, but operated at a sweet spot making the second-order spectral density $\mathcal{S}^{(2)}(\omega)$ the leading contribution. (iii) When one or multiple two-level fluctuators change the noise spectral density in the sub-Hz regime. (b) Nomenclature of the decay rates discussed in this work. The subscript "a" always indicates a decay into the antenna. (c) Power spectral density of thermal fields, coherent states, and shot noise plotted versus frequency. All fields can be filtered by the Lorentzian filter function of a resonator. The dashed line shows the transition frequency of the qubit.}
\end{figure*}

In this work, we systematically study the effect of the second-order coupling between thermal fields generated by a black-body radiator and a superconducting transmon qubit~\cite{Koch_2007} placed in a superconducting resonator. The novel aspect of our experiments is that we can either irradiate the qubit directly or via the resonator filter function with thermal noise of controllable power spectral density while keeping the qubit at the base temperature of a dilution refrigerator. This allows us to quantify the impact of thermal noise without suffering from parasitic effects such as quasiparticle generation in the superconducting circuits. Our work establishes thermal fields as a convenient tool to probe the coherence properties of quantum circuits in the microwave regime. In this way, we gain important insight into second-order decoherence mechanisms of superconducting qubits. Furthermore, our quantitative analysis of the decoherence rates is crucial to optimize the performance of superconducting qubits in a thermal environment.

After discussing the experimental setup in Sec.\,\ref{sec:exp}, we characterize the qubit coherence properties in the absence of thermal fields and, in addition, discuss first-order effects (see Sec.\,\ref{sec:intrinsic}). In the following sections, we analyze the three individual decoherence mechanisms depicted in Fig.\,\ref{fig:scheme}\,(a). We start with a discussion of energy relaxation of the qubit due to dispersively coupled thermal noise in Sec.\,\ref{sec:relaxation} [case~(i) in Fig.\,\ref{fig:scheme}\,(a)]. In the dispersive Jaynes-Cummings regime, noise at the resonator frequency couples in second-order. In our experiments, we find a coupling to broadband fields, which is enhanced as compared to that expected from the Purcell filter effect of the resonator. Furthermore, using coherent states and narrow-band shot noise, we demonstrate the counter-intuitive effect that the qubit relaxation rate is decreasing for increasing field strengths. In Sec.\,\ref{sec:secondorder}, we discuss the situation when the thermal noise field is directly irradiated on the qubit via a near-field antenna without the cavity filter [case~(ii) in Fig.\,\ref{fig:scheme}\,(a)]. At the flux sweet spot, this direct irradiation reveals the influence of second-order-coupled noise on the qubit dephasing rate. In particular, we observe the expected~\cite{Shnirman_2002,Makhlin_2004} $T^{3}$ temperature dependence of the qubit dephasing rate. In Sec.\,\ref{sec:longtime}, we show that low-frequency fluctuations of the qubit relaxation rate are related to the temperature of the black-body radiator if the field is not Purcell filtered [case~(iii) in Fig.\,\ref{fig:scheme}\,(a)]. We can explain this effect by the presence of two-level fluctuators in the spatial vicinity of the qubit, which change the effective noise spectral density. We show that it is crucial to apply noise-filtering techniques to the antenna line, which couples a broadband thermal frequency spectrum to the qubit. Finally, we conclude this article in Sec.\,\ref{sec:summary} with a comparison of the different decoherence mechanisms, which are summarized in Fig.\,\ref{fig:scheme}\,(b) and in Tab.\,\ref{tab:tab2}.

\section{Experimental Techniques}
\label{sec:exp}

The experimental setup to study the effect of thermal noise on a superconducting transmon qubit is sketched in Fig.\,\ref{fig:scheme}\,(a). In two different cooldowns, we couple a thermal noise source either directly through a near-field antenna or indirectly via a coplanar waveguide resonator to a transmon qubit. In our experiments, we measure the qubit coherence properties as a function of the power emitted by the noise source. We do not observe significant changes in the coherence properties of the sample between the two cooldowns. The noise source generating the propagating thermal fields at low temperatures is a \SI{30}{\decibel} attenuator, which is thermally decoupled from the sample box. Because the attenuator is also only weakly coupled to the base temperature stage of a dilution refrigerator, we can heat it up to $T\,{=}\,\SI{1.5}{\kelvin}$ for the emission of black-body radiation. We use a microwave attenuator with a frequency-independent admittance $Y(\omega)^{-1}\,{=}\,Z_{0}\,{=}\,\SI{50}{\ohm}$. In a short circuit configuration, such an attenuator would dissipate the power $\mathcal{P}_{\mathrm{sc}}\,{=}\,\mathcal{S}_{\mathrm{sc}}(\omega,T)\delta f$ in a frequency interval $\delta f$. Here, the noise power spectral density~\cite{Martinis_2003} $\mathcal{S}_{\mathrm{sc}}(\omega,T)\,{=}\,2\hbar\omega\coth(\hbar\omega/2k_{\mathrm{B}}T)$ [units \SI{}{\watt\per\hertz}] defines the frequency distribution of the power fluctuations dissipated in the attenuator. In our setup, we do not operate the attenuator in a short circuit configuration but couple it to a transmission line with an impedance of \SI{50}{\ohm}. Hence, the noise power spectral density $\mathcal{S}(\omega,T)$ propagating into the transmission line is reduced by a factor 4~\cite{Pozar_2012,Mariantoni_2010}, which yields
\begin{equation}
   \mathcal{S}(\omega,T) = \mathcal{S}_{\mathrm{sc}}(\omega,T)/4 = \hbar\omega(n_{\mathrm{th}}+1/2)\,.\label{eqn:SV2}
\end{equation}
Here, the average number of emitted thermal noise photons $n_{\mathrm{th}}(\omega,T)\,{=}\,[\exp(\hbar\omega/k_{\mathrm{B}}T)\,{-}\,1)]^{-1}$ is given by the Bose-Einstein distribution~\cite{Bose_1924}.

The qubit is located inside a sample box, which is mounted to the base temperature stage and stabilized to $T_{\mathrm{i}}\,{=}\,\SI{35}{\milli\kelvin}$. We operate the qubit in the dispersive regime, where the detuning~$\delta\,{\equiv}\,\omega_{\mathrm{q},0}\,{-}\,\omega_{\mathrm{r}}$ fulfills $g/\delta\,{\ll}\,1$ (see Tab.\,\ref{tab:tab1} for parameter values). Here, $g$ is the qubit-resonator coupling strength, $\omega_{\mathrm{q},0}$ is the qubit transition frequency at the flux sweet spot, and $\omega_{\mathrm{r}}$ is the resonator frequency. The resonator is further characterized by its internal and external quality factors $Q_{\mathrm{i}}\,{=}\,\omega_{\mathrm{r}}/\kappa_{\mathrm{i}}\,{=}\,\num{1.2e5}$ and $Q_{\mathrm{x}}\,{=}\,\omega_{\mathrm{r}}/\kappa_{\mathrm{x}}\,{=}\,\num{714}$, respectively. For a transmon qubit, which is not a perfect two-level system, the dispersive shift~\cite{Koch_2007} $\chi\,{\equiv}\,{-}g^{2}E_{\mathrm{c}}/[\delta(\delta\,{-}\,E_{\mathrm{c}})]$ depends on the transmon charging energy $E_{\mathrm{c}}$ and can be used for readout~\cite{Schuster_2005} and photon number calibration (see App.\,\ref{sec:AppSetup} for details).

\begin{table}[b]
\caption{\label{tab:tab1} Relevant parameter values for the characterized sample. Qubit decoherence rates are summarized in Tab.\,\ref{tab:tab2}.}
\begin{ruledtabular}
\begin{tabular}{lcc}
qubit transition frequency & $\omega_{\mathrm{q},0}$ & $2\pi\,{\times}\,\SI{6.92}{\giga\hertz}$ \\
qubit charging energy & $E_{\mathrm{c}}$ & $2\pi\,{\times}\,\SI{315}{\giga\hertz}$ \\
resonator frequency & $\omega_{\mathrm{r}}$ & $2\pi\,{\times}\,\SI{6.07}{\giga\hertz}$ \\
qubit-resonator detuning & $\delta$ & $2\pi\,{\times}\,\SI{850}{\mega\hertz}$ \\
qubit-resonator coupling & $g$ & $2\pi\,{\times}\,\SI{67}{\mega\hertz}$ \\
dispersive shift & $\chi$ & $-2\pi\,{\times}\,\SI{3.11}{\mega\hertz}$ \\
external resonator coupling & $\kappa_{\mathrm{x}}$ & $2\pi\,{\times}\,\SI{8.5}{\mega\hertz}$ \\
internal resonator loss & $\kappa_{\mathrm{i}}$ & $2\pi\,{\times}\,\SI{50}{\kilo\hertz}$ \\
resonator-antenna coupling & $\kappa_{\mathrm{a}}$ & $2\pi\,{\times}\,\SI{30}{\kilo\hertz}$ 
\end{tabular}
\end{ruledtabular}
\end{table}

\begin{figure}[t]
\includegraphics{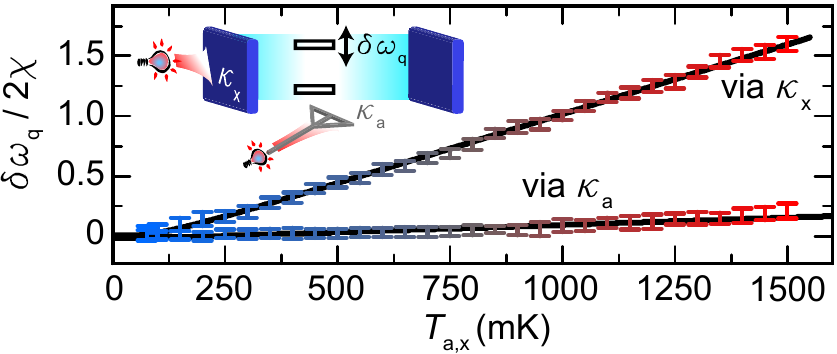}
\caption{\label{fig:corr} ac-Stark shift of the qubit transition frequency plotted versus temperature of the heatable attenuator in the readout line (via $\kappa_{\mathrm{x}}$) or the antenna line (via $\kappa_{\mathrm{a}}$). Solid lines are fits using Eq.\,(\ref{eqn:acstark1}) and  Eq.\,(\ref{eqn:acstark2}) while the inset depicts the experimental configuration.}
\end{figure}

We calibrate the thermal noise power on the input lines using the qubit as a power detector for the mean photon number $n_{\mathrm{r}}$ in the resonator. Since we generate the thermal states outside the resonator, the resonator population is calculated as a cavity field that is coupled to three bosonic baths defined in the following by the subscript $j\,{\in}\,\{\text{i},\text{x},\text{a}\}$. The first bath is the direct sample environment with a thermal occupation number $n_{\mathrm{i}}$ coupling via the internal loss rate $\kappa_{\mathrm{i}}$ to the resonator. Second, the resonator is coupled with external coupling rate $\kappa_{\mathrm{x}}$ to modes on the readout line, which can be thermally occupied by heating the attenuator, thereby generating the occupation number $n_{\mathrm{x}}$. In the same manner, we control the number of noise photons $n_{\mathrm{a}}$ on the antenna line coupling with rate $\kappa_{\mathrm{a}}$ to the resonator. The three coupling rates cause a resonator population $n_{\mathrm{r}}(\omega,T)\,{\approx}\,\mathcal{F}_{\mathrm{L}}(\omega)\sum_{j}\kappa_{j}n_{\mathrm{th},j}(\omega,T)$ (see  App.\,\ref{sec:broadband} for details). In the latter expression, $\mathcal{F}_{\mathrm{L}}(\omega)\,{=}\,(\kappa_{\mathrm{tot}}/2)/[(\kappa_{\mathrm{tot}}/2)^{2}\,{+}\,(\omega\,{-}\,\omega_{\mathrm{r}})^{2}]$ is the Lorentzian filter function of the resonator shown in Fig.\,\ref{fig:scheme}\,(c) and $\kappa_{\mathrm{tot}}\,{=}\,\kappa_{\mathrm{i}}\,{+}\,\kappa_{\mathrm{x}}\,{+}\,\kappa_{\mathrm{a}}$ is the total loss rate of the resonator. Because of the low sample temperature of \SI{35}{\milli\kelvin}, we can neglect thermal photons coupling via $\kappa_{\mathrm{i}}$. Then, on resonance, the steady state limit of the Markovian master equation [Eq.\,(\ref{eqn:markovres})] describing the resonator yields $n_{\mathrm{r}}(\omega_{\mathrm{r}},T_{\mathrm{x}},T_{\mathrm{a}})\,{=}\,[\alpha_{\mathrm{x}}\kappa_{\mathrm{x}}n_{\mathrm{x}}(\omega_{\mathrm{r}},T_{\mathrm{x}})\,{+}\,\alpha_{\mathrm{a}}\kappa_{\mathrm{a}}n_{\mathrm{a}}(\omega_{\mathrm{r}},T_{\mathrm{a}})]/\kappa_{\mathrm{tot}}$. Here, the factors $\alpha_{\mathrm{x}}$ and $\alpha_{\mathrm{a}}$ account for losses in the microwave lines between attenuator and sample. Because we use the very same combination of attenuator and coaxial cables for the two cool-downs, we assume $\alpha_{\mathrm{x}}\,{=}\,\alpha_{\mathrm{a}}\,{\equiv}\,\alpha$ in the following. In the dispersive regime, we calibrate this factor by measuring the ac-Stark shift~\cite{Schuster_2005} of the qubit
\begin{align}
 \delta\omega_{\mathrm{q,x}}(T_{\mathrm{x}}) &=  2\chi\alpha\kappa_{\mathrm{x}}\,\,n_{\mathrm{x}}(T_{\mathrm{x}})/\kappa_{\mathrm{tot}} + \delta\omega_{\mathrm{q,a}}(\SI{50}{\milli\kelvin})\,,
\label{eqn:acstark1}\\
 \delta\omega_{\mathrm{q,a}}(T_{\mathrm{a}}) &=  2\chi\alpha\kappa_{\mathrm{a}}n_{\mathrm{a}}(T_{\mathrm{a}})/\kappa_{\mathrm{tot}} + \delta\omega_{\mathrm{q,x}}(\SI{50}{\milli\kelvin})\,.\label{eqn:acstark2}
\end{align}
From sweeping the temperature $T_{\mathrm{x}}$ of the feedline attenuator, we obtain $\alpha\,{\simeq}\,\SI{4.1}{\decibel}$ using a numerical fit of Eq.\,(\ref{eqn:acstark1}) as shown in Fig.\,\ref{fig:corr}. We attribute this loss mainly to impedance mismatches either of the attenuator itself or of the cryogenic connection between attenuator and resonator. From a temperature sweep of $T_{\mathrm{a}}$, we extract the negligibly small coupling rate $\kappa_{\mathrm{a}}/2\pi\,{\simeq}\,\SI{30}{\kilo\hertz}$. We note that we also use coherent states and shot noise with the spectral density shown in Fig.\,\ref{fig:scheme}\,(b) in our experiments. These fields are generated at room temperature with state-of-the-art microwave equipment as discussed in App.\,\ref{sec:AppSetup}. We calibrate the photon number of these fields with ac-Stark shift measurements similar to the ones discussed above.

\section{Intrinsic qubit coherence and first-order coupling}
\label{sec:intrinsic}

\begin{figure}[t]
\includegraphics{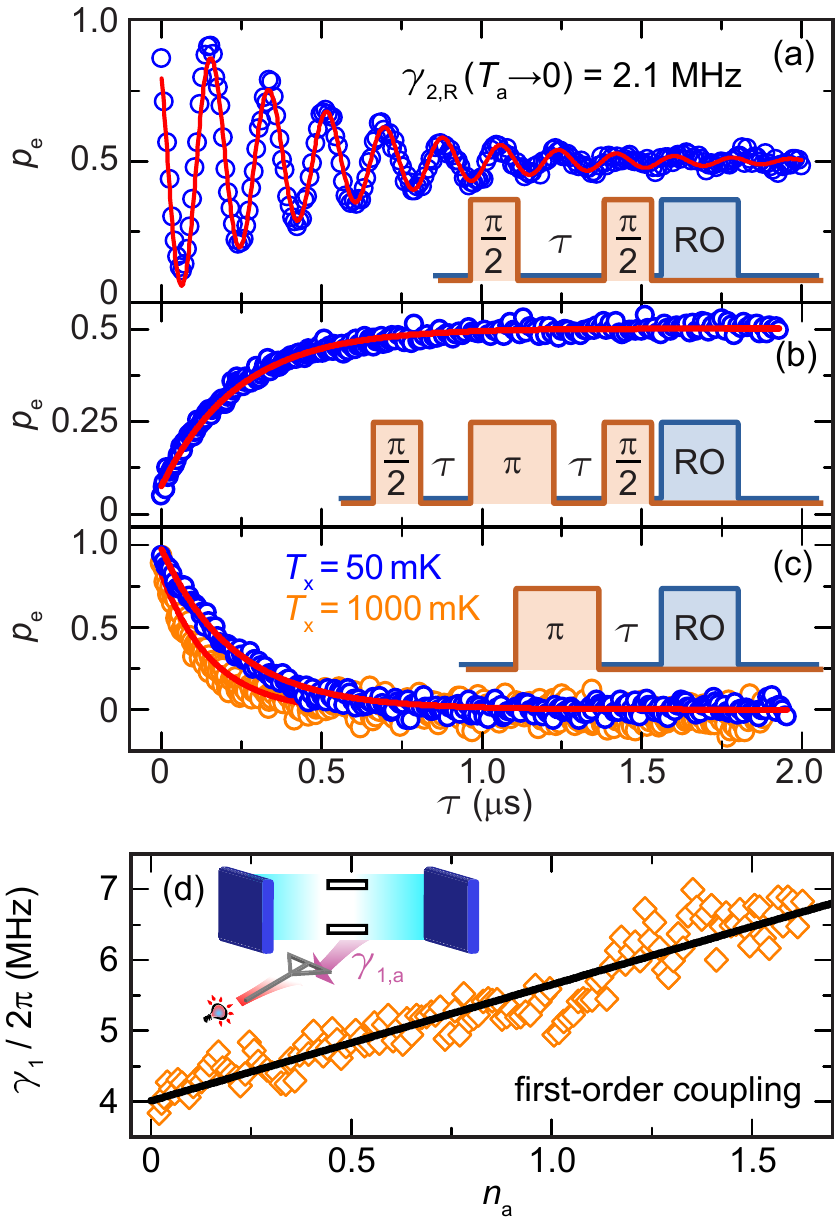}
\caption{\label{fig:T2}(a) Probability $p_{\mathrm{e}}$ to find the qubit in the excited state as a function of the free evolution time for the Ramsey pulse sequence (inset). Red envelopes indicate qubit control pulses, blue envelopes the readout pulse to the resonator. The solid lines are fits to the data points (circles). (b) As in panel (a), but for the spin-echo sequence. (c) As in panel (a), but for the relaxation sequence. $T_{\mathrm{x}}$ indicates the temperature of the readout line attenuator. (d) Qubit relaxation rate plotted versus the number of thermal photons emitted from the antenna line attenuator. The solid line is a linear fit and the inset depicts a sketch of the experimental setup.}
\end{figure}

In the absence of external microwave fields and at the flux sweet spot, the qubit is relaxation-limited with average coherence times of approximately \SI{500}{\nano\second}. In particular, we find typical values of the Ramsey decay rate $\gamma_{2,\mathrm{R}}/2\pi\,{\simeq}\,\SI{2.1}{\mega\hertz}$, the spin-echo decay rate $\gamma_{2,\mathrm{se}}/2\pi\,{\simeq}\,\SI{1.9}{\mega\hertz}$, and the temperature-independent relaxation rate $\gamma_{1}^{0}/2\pi\,{\simeq}\,\SI{3.9}{\mega\hertz}$ [cf.~Fig.\,\ref{fig:T2}\,(a)\,--\,(c)]. This relaxation rate is a factor of ten larger than the expected Purcell rate and most likely dominated by loss into the Si/SiO$_2$ substrate and into the on-chip antenna. The above numbers imply a pure dephasing rate $\gamma_{\varphi}^{0}\,{=}\gamma_{2,\mathrm{R}}\,{-}\,\gamma_{1}^{0}/2\,{\simeq}\,2\pi\,{\times}\,\SI{150}{\kilo\hertz}$.

We irradiate thermal states through the antenna on the qubit to measure its first-order decay rate $\gamma_{1\mathrm{,a}}$ into the antenna line. We assume that the thermal noise is Gaussian and weak $[\mathcal{S}(\omega)\,{\ll}\,\hbar\omega]$ for average photon numbers $n_{\mathrm{a}}\,{\lesssim}\,1$. Hence, we can apply the spin-boson model and Fermi's golden rule to obtain $\gamma_{1}\,{\propto}\,\mathcal{S}(\omega_{\mathrm{q}})/2\hbar$. This relation has been widely used to measure the frequency dependence $\mathcal{S}(\omega)$ of different noise sources by tuning the qubit transition frequency~\cite{Astafiev_2004,Houck_2008,Reed_2010,Kim_2011,Bylander_2011,Zaretskey_2013,Bronn_2015,Haeberlein_2015}. Here, we use a complementary approach and vary the magnitude of $\mathcal{S}(\omega_{\mathrm{q}}\,{=}\,\mathrm{const.})$ by controlling the number of noise photons $n_{\mathrm{a}}\,{\propto}\,\mathcal{S}(\omega_{\mathrm{q}})$. We expect $\gamma_{1}$ to increase linearly with $\mathcal{S}(\omega_{\mathrm{q}})$ and therefore with $n_{\mathrm{a}}$, which yields
\begin{equation}
\gamma_{1}(n_{\mathrm{a}}) = \tilde{\gamma}_{1}^{0} + \gamma_{1\mathrm{,a}}[2n_{\mathrm{a}} + 1]\,.
\label{eqn:deltagamma1}
\end{equation}
Here, $\tilde{\gamma}_{1}^{0}\,{=}\,\gamma_{1}^{0}\,{-}\,\gamma_{1\mathrm{,a}}$ is the temperature-independent relaxation rate corrected for the influence of vacuum fluctuations on the antenna line. As shown in  Fig.\,\ref{fig:T2}\,(d), we obtain the qubit decay rate into the antenna line, $\gamma_{1\mathrm{,a}}/2\pi\,{=}\,\SI{820}{\kilo\hertz}$, from a linear fit of Eq.\,(\ref{eqn:deltagamma1}) to the data. This value implies that approximately \SI{20}{\percent} of the total qubit relaxation rate can be attributed to decay into the antenna. The accuracy of the qubit acting as noise spectrometer is limited by the standard deviation $\sigma/2\pi\,{\simeq}\,\SI{215}{\kilo\hertz}$ of the data obtained from the fit. This scatter can be attributed to additional low-frequency fluctuations of the relaxation rate discussed in detail in Sec.\,\ref{sec:longtime}. In general, the broadband thermal radiation also includes frequency components near the gap frequency of Al, $\Delta_{0}/h\,{\simeq}\,\SI{80}{\giga\hertz}$, which can introduce quasiparticle-induced decoherence~\cite{Lutchyn_2005,Schreier_2008,Martinis_2009,Catelani_2011,Catelani_2012,Sun_2012,Zanker_2015,Gustavsson_2016}. A significant effect of quasiparticles, however, would lead to a decrease of the qubit relaxation rate for moderate temperatures~\cite{Martinis_2009}, which we do not observe in our experiments. The reason is that the sample itself is stabilized at \SI{35}{\milli\kelvin} and \SI{80}{\giga\hertz} radiation is strongly suppressed in the coaxial cables.

\section{Thermal fields in the dispersive Jaynes-Cummings regime}
\label{sec:relaxation}

\begin{figure}[t]
\includegraphics{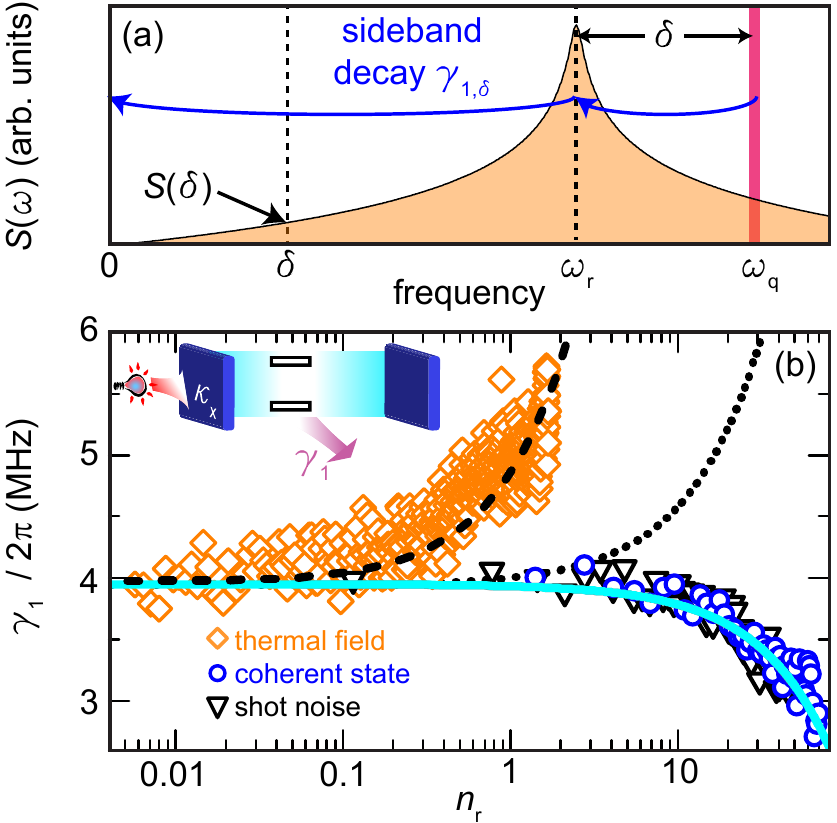}
\caption{\label{fig:gammarad} (a) Schematic drawing of sideband decay due to a finite noise power spectral density $\mathcal{S}(\delta)$ at the detuning frequency $\delta\,{=}\,\omega_{\mathrm{q}}\,{-}\,\omega_{\mathrm{r}}$. (b) Qubit relaxation rate for microwave fields coupling the qubit through the resonator (see inset for setup sketch) plotted versus the average photon number $n_{\mathrm{r}}$. The solid line is a numerical fit using Eq.\,(\ref{eqn:deltagammar}). The dotted line is a calculation based on Eq.\,(\ref{eqn:gammar}) modeling the expected increase due to thermal fields. For the dashed line we use a numerical fit of Eq.\,(\ref{eqn:gammar}), where the enhanced relaxation rate is modeled by $n_{\mathrm{q}}(T_{\mathrm{x}})$ as a free parameter.}
\end{figure}

In contrast to the direct irradiation of the qubit by thermal noise as discussed above, in this section we study the effect of thermal noise, which we apply through a superconducting resonator to the transmon qubit. The resonator acts as a narrow filter for the broadband thermal noise. Since qubit and resonator are far detuned, $(|\chi|\,{\ll}\,g)$, noise couples only in second-order, which is described within the dispersive limit of the Jaynes-Cummings model. There, the power spectral density $\mathcal{S}(\omega_{\mathrm{q}})$ at the qubit frequency is strongly reduced due to the Lorentzian filter function of the resonator. Nevertheless, in a second-order approximation~\cite{Blais_2004} also the noise spectral density $\mathcal{S}(\omega_{\mathrm{r}})$ at the resonator frequency couples dispersively to the qubit with the rate $\chi$. In an elaborate treatment, the broadband nature of thermal fields requires an exact transformation of the dispersive Jaynes-Cummings Hamiltonian, which takes sideband decay into account~\cite{Boissonneault_2009}. This transformation yields the total relaxation rate in the dispersive regime
\begin{align}
  \gamma_{1}(n_{\mathrm{r}},n_{\mathrm{q}}) = \gamma_{1}^{0} &+ \gamma_{1\mathrm{,P}}[2n_{\mathrm{q}}+1]\notag\\
& + [\gamma_{1,\delta} - \gamma_{\mathrm{mix}}][2n_{\mathrm{r}}+1] \,.
  \label{eqn:gammar}
\end{align}
For our specific sample, $\gamma_{1}^{0}$ is the main contribution to $\gamma_{1}$. Under thermal radiation, noise photons at the qubit frequency, $n_{\mathrm{q}}\,{\equiv}\,\alpha n_{\mathrm{th}}(\omega_{\mathrm{q}},T_{\mathrm{x}})$, enhance $\gamma_{1}$ by the Purcell decay rate~\cite{Purcell_1946,Sete_2014} $\gamma_{1\mathrm{,P}}\,{=}\,\kappa_{\mathrm{x}}g^{2}/\delta^{2}\,{\simeq}\,2\pi\,{\times}\,\SI{53}{\kilo\hertz}$. The third term in Eq.\,(\ref{eqn:gammar}) describes a competing mechanism between a reduced qubit decay with rate $\gamma_{\mathrm{mix}}\,{=}\,|\gamma_{1}^{0}\chi/\delta|\,{\simeq}\,2\pi\,{\times}\,\SI{14.3}{\kilo\hertz}$ due to the mixing of qubit and resonator states~\cite{Boissonneault_2009} and an enhancement with rate $\gamma_{1,\delta}\,{=}\,\hbar^{-1}|4\chi\mathcal{S}(\delta)/\delta|$. The latter decay rate reflects a sideband process resulting from the combined action of resonator photons at $\omega_{\mathrm{r}}$ and a thermal noise power $\mathcal{S}(\delta)$ as shown in Fig.\,\ref{fig:gammarad}\,(a). Interestingly, a reduction of the total decay rate $\gamma_{1}$ can be obtained for increasing $n_{\mathrm{r}}$ if $\gamma_{\mathrm{mix}}\,{>}\,\gamma_{1,\delta}$. We calibrate the rate $\gamma_{1,\delta}$ by selectively driving at the resonator frequency, which changes $n_{\mathrm{r}}$ while leaving $n_{\mathrm{q}}$ and $\gamma_{1,\delta}$ in Eq.\,(\ref{eqn:gammar}) constant. To this end, we use a coherent drive at the resonator frequency and measure the change of the relaxation rate
\begin{equation}
\delta\gamma_{1,\mathrm{r}}(n_{\mathrm{r}}) \equiv 2n_{\mathrm{r}}[\gamma_{1,\delta} - \gamma_{\mathrm{mix}}]\,,
\label{eqn:deltagammar}
\end{equation}
which is obtained by keeping only the $n_{\mathrm{r}}$-dependent terms on the right hand side of Eq.\,(\ref{eqn:gammar}). As shown in Fig.\,\ref{fig:gammarad}\,(b), we observe a decrease of the qubit relaxation rate for a coherent drive with $\delta\gamma_{1,\mathrm{r}}(n_{\mathrm{r}})/2\pi\,{=}\,{-}\SI{17}{\kilo\hertz}$/photon obtained from a numerical fit based on Eq.\,(\ref{eqn:deltagammar}). This decrease of relaxation rate yields a sideband decay rate $\gamma_{1,\delta}/2\pi\,{\simeq}\,\SI{5.8}{\kilo\hertz}$ equivalent to $\mathcal{S}(\delta)\,{\simeq}\,\SI{2.6e-28}{\watt\per\hertz}$. We obtain the same result when irradiating the resonator with shot noise that has a spectral density as shown in Fig.\,\ref{fig:scheme}\,(b). Please note that the overall qubit decoherence rate given by $\gamma_{2}\,{=}\,\gamma_{\varphi}+\gamma_{1}/2$ is nevertheless increasing due to the additional dephasing from noise in the photon number as discussed in detail in Ref.~\onlinecite{Goetz_2016a}.

When we apply broadband thermal fields through the resonator input, we experimentally verify the three competing decay rates $\gamma_{1\mathrm{,P}},\,\gamma_{1,\delta}$, and $\,{-}\gamma_{\mathrm{mix}}$ present in Eq.\,(\ref{eqn:gammar}). Because $\gamma_{1\mathrm{,P}}\,{+}\,\gamma_{1,\delta}\,{>}\,\gamma_{\mathrm{mix}}$, the overall relaxation rate increases with increasing number of thermal photons stored inside the resonator as shown in Fig.\,\ref{fig:gammarad}\,(b). When comparing the measured total relaxation rate $\gamma_{1}$ with a calculation according to Eq.\,(\ref{eqn:gammar}) [dotted line in Fig.\,\ref{fig:gammarad}\,(b)], we find that the measured decay rate is significantly larger than expected. Hence, we fit Eq.\,(\ref{eqn:gammar}) to the data and use $n_{\mathrm{q}}(T_{\mathrm{x}})$ as a free parameter [dashed line in Fig.\,\ref{fig:gammarad}\,(b)]. We find that the coupling of the thermal noise fields to the qubit is enhanced by a factor of $10$. Because the qubit is galvanically decoupled from the resonator and the sample is stabilized at \SI{35}{\milli\kelvin}, this additional coupling is most likely mediated by parasitic modes of the sample box~\cite{Houck_2008,Schuster_2001}. Among others, such modes can be slotline, parallel plane, and surface wave modes with resonance frequencies close to the qubit transition frequency. Our results show that this mechanism originating from the broadband nature of the noise fields can dominate over the associated Purcell rate originating from the finite bandwidth of the resonator. Therefore, great care must be taken in the microwave design of sample holders and chip layout in order to minimize losses from broadband fields.

\section{Second-order flux noise from thermal fields}
\label{sec:secondorder}

In addition to the relaxation processes discussed above, the qubit can suffer from dephasing due to propagating thermal fields, even if the resonator filters them. The main contribution in this context is noise, which introduces dephasing by modulating the qubit frequency via the ac-Stark shift~\cite{Bertet_2005a,Schuster_2005,Sears_2012}. We analyze this effect for thermal fields on the readout line in detail in Ref.~\onlinecite{Goetz_2016a}. Here, we focus on dephasing caused by second-order intensity fluctuations of the thermal fields emitted from the heatable attenuator in the antenna line. The thermal noise fields manifest as current fluctuations on the short-circuited on-chip antenna, which couple magnetic flux noise into the SQUID loop of the transmon qubit. Far away from the flux sweet spot, the qubit dephasing is dominated by the first-order noise power spectral density $\mathcal{S}(\omega{\mapsto}0)$ defined in Eq.\,(\ref{eqn:SV2}) due to a finite first-order transfer function (see App.\,\ref{sec:2ndordercoupling} for details). At the flux sweet spot, however, the first-order transfer function vanishes while second-order intensity fluctuations can still introduce dephasing. These fluctuations are characterized by the second-order power spectral density~\cite{Shnirman_2002,Makhlin_2004} (units \SI{}{\square\watt\per\hertz})
\begin{equation}
\mathcal{S}^{(2)}(\omega) = \omega\left[\frac{\hbar^{2}\omega^{2}+4\pi^{2}k_{\mathrm{B}}^{2}T^{2}}{12\pi}\right]\coth\left(\frac{\hbar\omega}{2k_{\mathrm{B}}T}\right)
\label{eqn:Sth2}
\end{equation}
Using the second-order transfer function $\mathcal{D}^{(2)}_{\lambda,\mathrm{z}}\,{=}\,\pi^{2}\omega_{\mathrm{q},0}/2$ at the sweet spot, we find the thermally induced qubit dephasing rate [cf.\,Eq.\,(\ref{eqn:gammaphi2})]
\begin{figure}[t]
\includegraphics{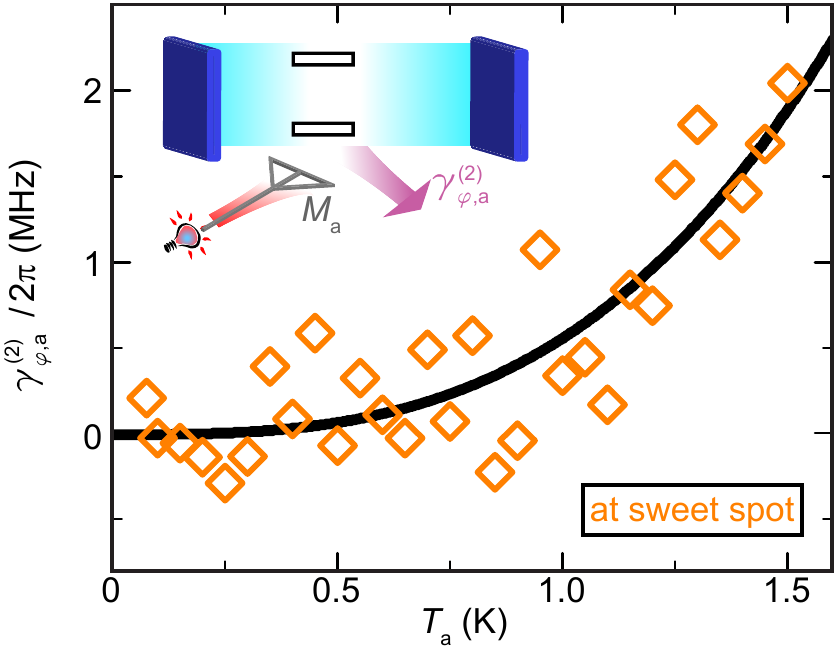}
\caption{\label{fig:2ndorder} Thermally induced qubit dephasing rate $\gamma_{\varphi\mathrm{,a}}^{(2)}(T_{\mathrm{a}})$ measured as a function of the temperature $T_{\mathrm{a}}$ of the black-body radiator as depicted in the inset. To isolate second-order effects, we operate the qubit at the flux sweet spot of the transmon qubit. The solid line shows the $T_{\mathrm{a}}^{3}$ dependence obtained from a numerical fit using Eq.\,(\ref{eqn:gphi_2nd}).}
\end{figure}
\begin{equation}
 \gamma_{\varphi\mathrm{,a}}^{(2)}(T_{\mathrm{a}}) = 2\pi\left[\frac{\pi^{2}}{4\sqrt{3}}\frac{M_{\mathrm{a}}^{2}}{L_{\ell}Z_{0}}\right]^{2} \left[\frac{k_{\mathrm{B}}T_{\mathrm{a}}}{\hbar}\right]^{3}\,,
\label{eqn:gphi_2nd}
\end{equation}
which follows a $T_{\mathrm{a}}^{3}$ dependence. Here, $M_{\mathrm{a}}$ describes the mutual inductance between qubit and antenna, $L_{\ell}$ is the inductance of the SQUID loop, and $Z_{0}$ is the line impedance. In Fig.\,\ref{fig:2ndorder}, we plot the qubit dephasing rate $\gamma_{\varphi\mathrm{,a}}^{(2)}(T_{\mathrm{a}})\,{=}\,\gamma_{2,\mathrm{R}}(T_{\mathrm{a}})\,{-}\,\gamma_{2,\mathrm{R}}(T_{\mathrm{a}}{\mapsto}0)\,{-}\,\gamma_{1\mathrm{,a}}(T_{\mathrm{a}})$ caused by the intensity fluctuations of the thermal noise field at the flux sweet spot of the transmon qubit. The temperature independent decay rate $\gamma_{2,\mathrm{R}}(T_{\mathrm{a}}{\mapsto}0)$ and the thermally induced relaxation rate $\gamma_{1\mathrm{,a}}(T_{\mathrm{a}})$ are discussed in Sec.\,\ref{sec:intrinsic} [see Fig.\,\ref{fig:T2}\,(a) and Fig.\,\ref{fig:T2}\,(d), respectively]. The additional dephasing rate follows the expected $T_{\mathrm{a}}^{3}$ dependence with a scatter that is dominated by additional low-frequency fluctuations of the decay rate discussed in Sec.\,\ref{sec:longtime}. Fitting Eq.\,(\ref{eqn:gphi_2nd}) to the data, we find a loop inductance $L_{\ell}\,{\simeq}\,\SI{50}{\pico\henry}$ which is in reasonable agreement with the value of \SI{100}{\pico\henry} estimated from the loop geometry. Here, we have used the mutual inductance $M_{\mathrm{a}}\,{=}\,\SI{1.3}{\pico\henry}$ obtained from measuring the induced flux shift of the qubit when applying a DC current through the antenna line. Our results show that second-order flux noise can induce residual dephasing in a transmon qubit, even if it is operated at the flux sweet spot. For typical temperatures $(T_{\mathrm{a}}\,{<}\,\SI{50}{\milli\kelvin})$ used in circuit QED experiments, however, we find that additional intensity fluctuations introduce only negligible dephasing of approximately $\SI{100}{\hertz}$ because of the quadratic suppression $(k_{\mathrm{B}}T_{\mathrm{a}}/\hbar\omega_{\mathrm{q},0})^{2}\,{\ll}\,1$ [cf.\,Eq.\,(\ref{eqn:gammaphi2})]. Nevertheless, dephasing is not only determined by the relatively weak thermal contribution but also by stronger $1/f$ noise~\cite{Yoshihara_2006,Bylander_2011}. Because $1/f$ noise also has a second-order contribution~\cite{Makhlin_2004}, our results show that this noise can be a possible source for the residual dephasing found for transmon qubits. We do not find any indications for quasiparticle-induced dephasing mechanisms in our dephasing measurements. In particular, we do not observe the characteristic decay law expected for quasiparticle-induced dephasing~\cite{Zanker_2015}, $\gamma_{\varphi}\,{\propto}\,\exp[{-}x(t)]$, with $x(t)\,{\propto}\,t^{3/2}$. The reason is that the transmon qubit operates in a regime $E_{\mathrm{J}}\,{>}\,E_{\mathrm{c}}$ where quasiparticles have only negligible influence~\cite{Lutchyn_2006,Koch_2007,Schreier_2008} if the split junction has no significant asymmetry~\cite{Catelani_2012}.

\section{Fluctuating qubit parameters in the presence of thermal fields}
\label{sec:longtime}

In the following, we study the effect of thermal fields on the frequency spectrum of fluctuating qubit parameters. In particular, we analyze low-frequency variations of the qubit relaxation rate. This phenomenon was also observed for flux qubits~\cite{Yan_2012,Yan_2015,Gustavsson_2016}, for transmon qubits in a 3D cavity and phase qubits~\cite{Muller_2015}, as well as for the resonance frequency of superconducting resonators~\cite{Burnett_2014}. One prominent noise source in microscopic systems are fluctuating TLSs~\cite{Schickfus_1977,Strom_1978,Phillips_1987} hopping between two bistable spatial configurations. In superconducting circuits, the TLSs can for example be present in the thin oxide layer of the junction itself~\cite{Simmonds_2004,Shalibo_2010,Lisenfeld_2010} or in the dielectric environment of the qubit~\cite{Martinis_2005}. In these scenarios, the TLSs can be responsible for the low-frequency fluctuations of qubit parameters~\cite{Faoro_2015,Muller_2015}.

\begin{figure}[t]
\includegraphics{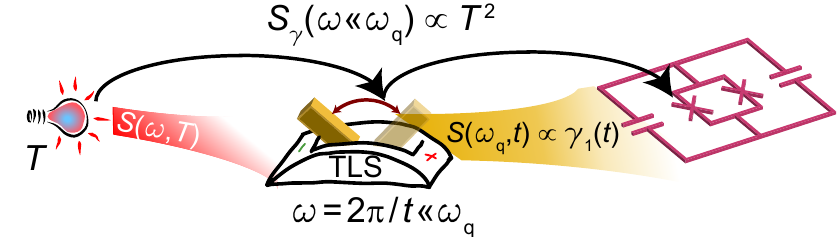}
\caption{\label{fig:spectral} TLS-mediated relaxation rate fluctuations. A thermal field influences the fluctuation rate of TLSs, which causes fluctuations of the noise power spectral density $\mathcal{S}(\omega_{\mathrm{q}},t)$.}
\end{figure}

Each individual TLS provides a Lorentzian shaped noise spectral density, which is centered around its excitation frequency $\omega_{\mathrm{tls}}\,{=}\,\sqrt{\varepsilon^{2}\,{+}\,\Delta^{2}}/\hbar$. Here, $\varepsilon$ is the asymmetry energy and $\Delta$ is the tunnel splitting. Because the TLSs are coupled to each other via dipole or strain-mediated interaction~\cite{Lisenfeld_2015}, each TLS eigenfrequency depends on the state of the other TLSs. The low-frequency variations of the TLS configuration results in low-frequency fluctuations of the noise power spectral density $\mathcal{S}(\omega_{\mathrm{q}},t)$ generated by the TLSs. Consequently, the qubit relaxation rate $\gamma_{1}(t)\,{=}\,\mathcal{S}(\omega_{\mathrm{q}},t)/2\hbar$ starts fluctuating. The rate of these fluctuations is influenced by a thermal field as depicted in Fig.\,\ref{fig:spectral}. Since qubit parameter fluctuations are typically recorded in the sub-Hz regime where $\hbar\omega\,{\ll}\,k_{\mathrm{B}}T$, we assume a white spectrum proportional to $k_{\mathrm{B}}T$ for the contribution of the environmental heat bath. For a distribution $P(\varepsilon,\Delta)d\varepsilon d\Delta\,{\propto}\,\varepsilon^{x}\Delta^{-1}d\varepsilon d\Delta$ of TLSs, we expect a $T^{2+x}$-dependence to be the dominant contribution to the spectrum of $\gamma_{1}$ \cite{Muller_2015}. This contribution arises from those TLSs, which are detuned by $\delta\omega\,{\equiv}\,|\omega_{\mathrm{q}}\,{-}\,\omega_{\mathrm{tls}}|\,{\gg}\,\gamma_{2\mathrm{,R}}$. The exponent $x\,{\geq}\,0$ is nonzero only for a finite interaction between the TLSs~\cite{Muller_2015}.

\begin{figure}[t]
\includegraphics{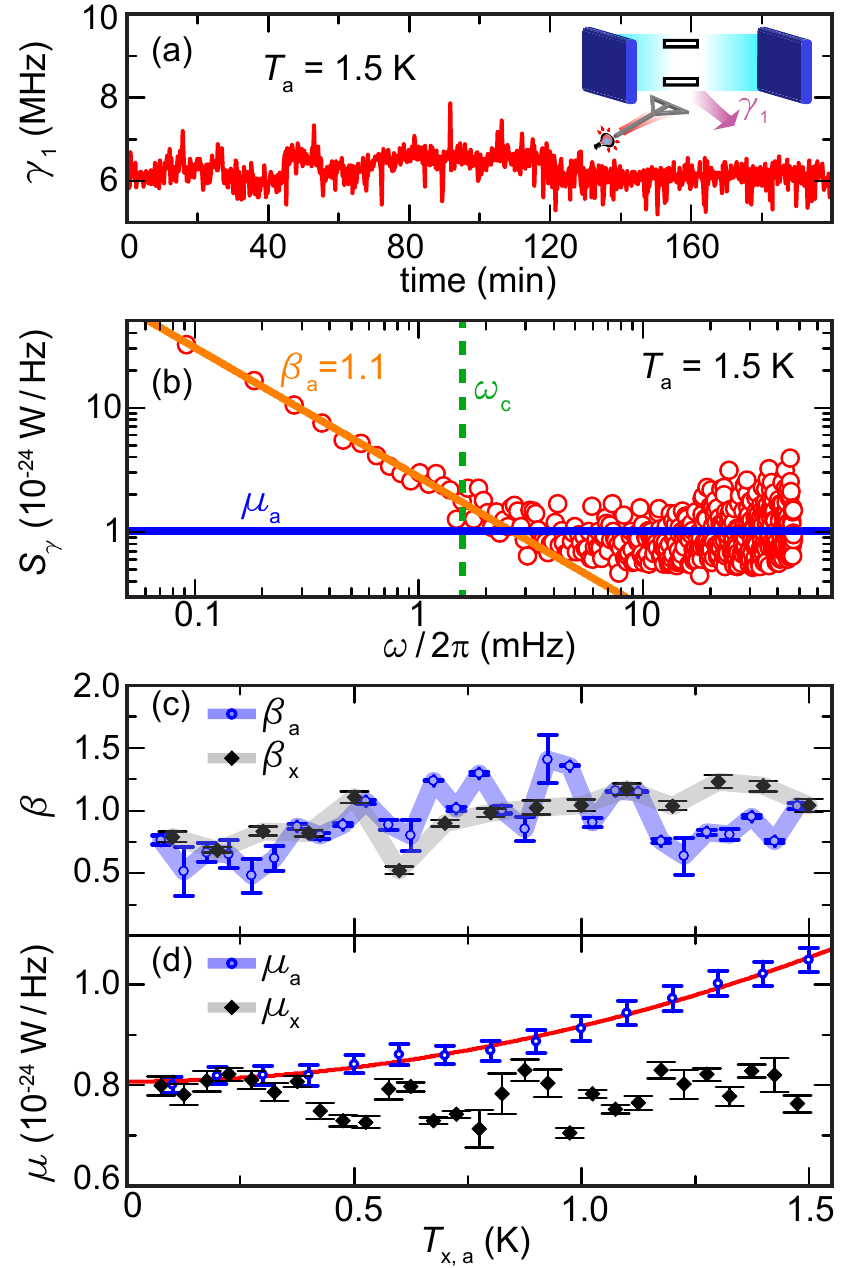}
\caption{\label{fig:Time_trace}(a) Qubit relaxation rate recorded as a function of time recorded over a period of \SI{200}{\minute} at $T_{\mathrm{a}}\,{=}\,\SI{1.5}{\kelvin}$. (b) Noise power spectral density for the data shown in panel (a). We obtain $\beta$ from a fit to the data with frequencies below $\omega_{\mathrm{c}}$ and $\mu$  as the mean value for points above $\omega_{\mathrm{c}}$. (c) Exponent $\beta$ measured for $\omega\,{<}\,\omega_{\mathrm{c}}$ versus temperature when applying thermal fields through the antenna or the resonator. Solid lines are guides to the eyes and error bars represent confidence intervals generated by the $\omega^{-\beta}$ fits. (d) Mean value $\mu$ of the white-noise contribution of thermal fields entering via the antenna or the resonator to $\mathcal{S}_{\gamma}$. The solid line is a fit as explained in text and error bars are the standard error of the mean.}
\end{figure}

We experimentally characterize the fluctuations of the noise power spectral density by measuring fluctuations of the qubit relaxation rate $\gamma_{1}(t)$. To this end, we perform systematic long-time measurements of the relaxation rate as a function of the temperatures $T_{\mathrm{x}}$ and of $T_{\mathrm{a}}$ ranging from \SI{50}{\milli\kelvin} to \SI{1500}{\milli\kelvin}. Each measurement is comprised of individual measurement traces taken at a repetition rate of $\SI{100}{\kilo\hertz}$, where the resonator is probed with \SI{2}{\micro\second} long traces with \SI{250}{\mega\hertz} sampling rate. We average \num{4e5} of these measurements to extract the decay rate at a particular moment and then wait for \SI{6}{\second} such that individual data points are recorded at a rate of approximately $\SI{0.1}{\hertz}$. A typical series of relaxation measurements at $T_{\mathrm{a}}\,{=}\,\SI{1500}{\milli\kelvin}$ over \SI{200}{\minute} is shown in Fig.\,\ref{fig:Time_trace}\,(a). Here, we observe a standard deviation $\sigma/2\pi\,{\simeq}\,\SI{320}{\kilo\hertz}$ from the mean relaxation rate $\langle\gamma_{1}\rangle/2\pi\,{\simeq}\,\SI{6.25}{\mega\hertz}$. Even though the absolute value of $\sigma$ seems large compared to other works~\cite{Muller_2015}, the relative scatter $\sigma/\langle\gamma_{1}\rangle\,{\simeq}\,0.05$ is comparable. In our experiments, we observe no systematic influence of the temperature on $\sigma$ for sweeps of $T_{\mathrm{a}}$ or  $T_{\mathrm{x}}$. To obtain more insight into the nature of the fluctuations, we investigate their spectral distribution. To this end, we calculate the Fourier transform $\mathcal{S}_{\gamma}(\omega)\,{=}\,\hbar/2\pi\int\mathrm{d}t\,\langle\gamma_{1}(t)\gamma_{1}(0)\rangle e^{-\imath\omega t}$ of the autocorrelation function $\langle\gamma_{1}(t)\gamma_{1}(0)\rangle$ as shown in Fig.\,\ref{fig:Time_trace}\,(b). For low frequencies, the data follows a $\omega^{-\beta}$-dependence, and crosses over into a frequency-independent tail for frequencies larger than a characteristic frequency $\omega_{\mathrm{c}}\,{\simeq}\,\SI{1}{\milli\hertz}$. As shown in Fig.\,\ref{fig:Time_trace}\,(c), we find that $\beta_{\mathrm{a}}\,{\simeq}\,0.91\,{\pm}\,0.24$ and  $\beta_{\mathrm{x}}\,{\simeq}\,0.96\,{\pm}\,0.19$ are approximately constant for thermal fields applied through antenna or resonator, respectively. This result is an extension of recent findings presented in Ref.\,\onlinecite{Yan_2012}, where $1/f$ fluctuations were analyzed up to a maximum temperature of \SI{200}{\milli\kelvin}.

Let us now turn to the white-noise contribution $\mu_{\mathrm{a}}$ above the characteristic frequency $\omega_{\mathrm{c}}$ exemplarily shown in Fig.\,\ref{fig:Time_trace}\,(b) and systematically plotted as a function of the antenna line attenuator in Fig.\,\ref{fig:Time_trace}\,(d). We observe an increase of $\mu_{\mathrm{a}}$ with the temperature $T_{\mathrm{a}}$. From a numerical fit based on the function $\mu_{\mathrm{a}}\,{=}\,\mu_{\mathrm{a0}}\,{+}\,aT_{\mathrm{a}}^{2+x}$, we find $\mu_{\mathrm{a0}}\,{=}\,\SI{0.81e-24}{\watt\per\hertz}$, $a\,{=}\,\SI{1.1e-25}{\watt\per\hertz\per\square\kelvin}$, and $x\,{=}\,{-}0.01\,{\pm}\,0.13$. Hence, our results support the model presented in Ref.~\onlinecite{Muller_2015} where a bath of TLSs acts as a source for the fluctuations in the qubit relaxation rate. In particular, the negligible value of $x\,{\simeq}\,0$ indicates that the TLSs relevant for our experiments are noninteracting. In contrast to $\mu_{\mathrm{a}}$, the white noise level $\mu_{\mathrm{x}}$ is approximately independent of the thermal field [see Fig.\,\ref{fig:Time_trace}\,(d)]. In this case, the resonator filters the thermal fields and protects the TLSs from the external noise. Our results show that a small resonator bandwidth and well-filtered feedlines are necessary in order to suppress externally activated switching of two-level fluctuators. Assuming that the fluctuation amplitude follows $\gamma_{1}$, also a smaller relaxation rate helps to minimize this effect.

\begin{table*}
\caption{\label{tab:tab2} Overview of the decoherence mechanisms discussed in this work and possible ways to improve them.}
\begin{ruledtabular}
\begin{tabular}{lccl}
\textbf{Relaxation rates} &  &  & possible improvement \\
\hline
Ramsey decay rate & $\gamma_{2\mathrm{,R}}$ & $2\pi\,{\times}\,\SI{2.1}{\mega\hertz}$ & better materials, better shielding \\
temperature-independent relaxation rate & $\gamma_{1}^{0}$ & $2\pi\,{\times}\,\SI{3.9}{\mega\hertz}$ & better materials, better shielding \\
contributions to $\gamma_{1}^{0}$: & & & \\
antenna line coupling & $\gamma_{1\mathrm{,a}}$ & $2\pi\,{\times}\,\SI{820}{\kilo\hertz}$ & smaller antenna coupling strength \\
Purcell decay rate & $\gamma_{1\mathrm{,P}}$ & $2\pi\,{\times}\,\SI{53}{\kilo\hertz}$ & smaller $\kappa_{\mathrm{x}}$ or smaller coupling $g$, larger detuning $\delta$ \\
sideband decay rate & $\gamma_{1,\delta}$ & $2\pi\,{\times}\,\SI{5.8}{\kilo\hertz}$ & smaller $\kappa_{\mathrm{x}}$ or smaller coupling $g$, larger detuning $\delta$ \\
qubit-resonator dressing effect & ${-}\gamma_{\mathrm{mix}}$ & ${-}2\pi\,{\times}\,\SI{14.3}{\kilo\hertz}$ &  \\
relaxation due to residual noise sources & $\gamma_{1}^{\mathrm{res}}$ & ${\sim}2\pi\,{\times}\,\SI{3}{\mega\hertz}$ & better substrate materials, better qubit materials \\
 & & & \\
\textbf{Dephasing rates} &  &  & possible improvement \\
\hline
pure dephasing rate & $\gamma_{\varphi}^{0}$ & $2\pi\,{\times}\,\SI{150}{\kilo\hertz}$ & better materials, better shielding \\
contributions to $\gamma_{\varphi}^{0}$: & & & \\
second-order antenna line coupling & $\gamma_{\varphi\mathrm{,a}}^{(2)}$ & $2\pi\,{\times}\,\SI{100}{\hertz}$ & smaller antenna coupling strength \\
photon shot noise [see Ref.\,\onlinecite{Goetz_2016a}] & $\gamma_{\varphi\mathrm{,n}}$ & $2\pi\,{\times}\,\SI{3.9}{\mega\hertz}/$photon &  larger $\kappa_{\mathrm{x}}$ or smaller dispersive shift $\chi$ \\
 & & & \\
\textbf{Qubit parameter fluctuations} &   & $T$-dependence & possible improvement \\
\hline
white resonator contribution & $\mu_{\mathrm{x}}$ & none & improved $\gamma_{1}$ \\
white resonator contribution & $\mu_{\mathrm{a}}$ & $T^{2}$ & smaller $\gamma_{1\mathrm{,a}}$ and improved $\gamma_{1}$ \\
\end{tabular}
\end{ruledtabular}
\end{table*}

\section{Summary and Conclusions}
\label{sec:summary}

In summary, we have characterized the influence of propagating thermal microwaves onto second-order decoherence mechanisms of a transmon qubit in a resonator. Because we spatially and thermally separate the thermal emitter from the circuit QED sample, we are able to separate the influence of the thermal noise from the residual loss channels of the qubit. This allows us to quantify three different second-order decoherence mechanisms. First, for the dispersive regime we find that the additional relaxation rate due to thermal fields applied via the resonator is larger than expected from Purcell filtering. This is a strong hint to the relevance of additional coupling channels such as parasitic on-chip modes. Second, we observe the expected $T^{3}$ dependence for the additional dephasing due to second-order noise at the flux sweet spot. This finding may explain the residual dephasing rates found for superconducting qubits with long coherence times. Finally, we investigate the influence of thermal fields on the low-frequency spectrum of qubit parameter fluctuations. We find that thermal fields enhance the white contribution of the noise power spectral density if applied broadband via an on-chip antenna. Our data confirms a model of thermally activated TLSs interacting with the qubit. The resonator, however, can filter this effect efficiently.

Finally, we compare the different decoherence mechanisms discussed in this work. A summary can be found in Tab.\,\ref{tab:tab2}. In our experiments, we perform relaxation measurements and measurements of the Ramsey decay rate $\gamma_{2\mathrm{,R}}/2\pi\,{\simeq}\,\SI{2.1}{\mega\hertz}$. For our sample and in absence of thermal fields, $\gamma_{2\mathrm{,R}}$ is dominated by the temperature-independent relaxation rate $\gamma_{1}^{0}/2\pi\,{\simeq}\,\SI{3.9}{\mega\hertz}$, which has several origins. In this work, we mainly discuss additional relaxation induced by thermal fields present on the microwave control lines. The sample geometry leads to a relatively large decay rate $\gamma_{1\mathrm{,a}}/2\pi\,{\simeq}\,\SI{820}{\kilo\hertz}$ into the antenna line, which can be improved by designing a smaller mutual inductance or capacitance between qubit and antenna. Furthermore, better filtering techniques can reduce noise on the antenna line. The second on-chip line coupling to the qubit is the microwave resonator, which is inevitable in most circuit-QED systems. However, the already small Purcell decay rate $\gamma_{1\mathrm{,P}}/2\pi\,{\simeq}\,\SI{53}{\kilo\hertz}$ could be further improved by using a resonator with a smaller bandwidth or a smaller coupling strength $g$ between qubit and resonator. These arguments also hold for the decay rate $\gamma_{1,\delta}/2\pi\,{\simeq}\,\SI{6.3}{\kilo\hertz}$ due to sideband processes. In addition to these decay processes into the microwave lines, our sample feels a strong contribution of residual noise sources leading to $\gamma_{1}^{\mathrm{res}}/2\pi\,{\simeq}\,\SI{3}{\mega\hertz}$. This decay rate is most likely dominated by loss into the Si/SiO$_{2}$ substrate and/or by quasiparticle generation from stray infrared light. Concerning dephasing, we show that second-order intensity fluctuations of thermal fields only have a negligible influence for our specific sample. The third noise mechanism discussed in this work, i.e., fluctuating qubit parameters, may have different origins~\cite{Burnett_2014,Faoro_2015,Muller_2015,Yan_2015,Gustavsson_2016}. Here, we show that thermally activated noninteracting TLSs are one possible reason for the observation of these fluctuations. The fact that the white noise spectrum of the fluctuations shows a weaker temperature dependence when the input field is Purcell-filtered, shows that this noise type can be efficiently filtered. This finding is especially relevant for quantum computation algorithm requiring a reliable operation over long timescales.

The authors acknowledge support from DFG through FE 1564/1-1, the doctorate program ExQM of the Elite Network of Bavaria, and the IMPRS \textquoteleft Quantum Science and Technology

\appendix

\begin{figure*}
\includegraphics{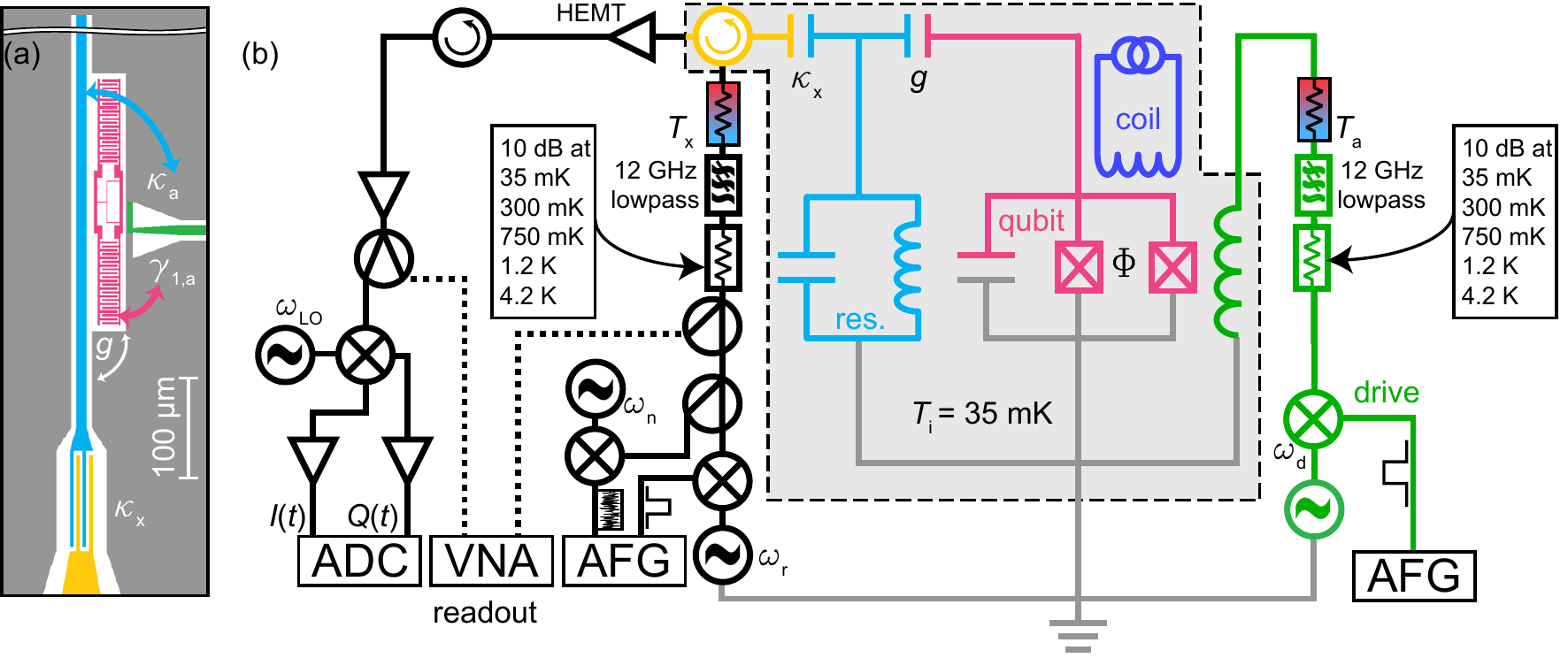}
\caption{\label{fig:setup}(a) Sample design: The frequency-tunable transmon qubit is capacitively coupled with coupling strength $g$ to a readout resonator, which itself is coupled with rate $\kappa_{\mathrm{x}}$ to a readout line. Furthermore, the qubit is coupled with rate $\gamma_{1\mathrm{,a}}$ to a $\SI{50}{\ohm}$-matched on-chip antenna and with rate $\kappa_{\mathrm{a}}$ to the resonator. (b) Schematic drawing of the experimental setup. We can inject thermal states into antenna and resonator by controlling the temperature $T_{\mathrm{x}}$ and $T_{\mathrm{a}}$ of heatable attenuators while stabilizing the sample stage (enclosed by the dashed box) at $T_{\mathrm{i}}\,{=}\,\SI{35}{\milli\kelvin}$.}
\end{figure*}

\section{Sample and measurement details}
\label{sec:AppSetup}

We use a superconducting transmon qubit coupled to a quarter wavelength coplanar waveguide resonator as depicted in Fig.\,\ref{fig:setup}\,(a). The qubit is characterized by the charging energy $E_{\mathrm{c}}\,{\simeq}\,h\,{\times}\,\SI{315}{\mega\hertz}$ and the total Josephson energy $E_{\mathrm{J},0}\,{\simeq}\,h\,{\times}\,\SI{20}{\giga\hertz}$ of the two SQUID junctions, which are used to tune the qubit transition frequency $(E_{\mathrm{J0}}/E_{\mathrm{c}}\,{\simeq}\,64)$. The qubit is made from a \SI{110}{\nano\meter} thick Al/AlO$_{\mathrm{x}}$/Al trilayer structure, shadow evaporated onto an undoped Si substrate. The silicon is covered with \SI{50}{\nano\meter} thermal oxide on either side and has a resistivity larger than \SI{1}{\kilo\ohm\centi\meter} at room temperature. We fabricate the \SI{50}{\ohm}-matched resonator with optical lithography from a \SI{100}{\nano\meter} thick Nb film. We mount the sample chip onto a copper-plated printed circuit board inside a gold-plated sample box made from copper.

To generate thermal states on the readout and on the antenna line, we use heatable \SI{30}{\decibel} attenuators integrated into the feedlines as depicted in Fig.\,\ref{fig:setup}\,(b). For the coaxial cables connecting the attenuators to the sample box, we use \SI{20}{\centi\meter} of Nb/CuNi UT47. The temperatures $T_{\mathrm{x,a}}$ of the heatable attenuators used to vary the thermal photon number can be individually controlled between \SI{0.050+-0.001}{\kelvin} and \SI{1.50+-0.01}{\kelvin}. Thermal noise from higher temperature stages has only a negligible influence for our setup due to individual \SI{10}{\decibel} attenuators in the feedlines at \SI{4}{\kelvin}, \SI{1.2}{\kelvin}, \SI{0.75}{\kelvin}, and \SI{0.3}{\kelvin}. We further use \SI{12}{\giga\hertz} low-pass filters on the two input lines, which are known to improve qubit coherence properties~\cite{Rigetti_2012,Geerlings_2013a,Stern_2014,Pop_2014}. In a configuration without these filters, the maximum Josephson energy $E_{\mathrm{J},0}$ of our qubit is reduced by \SI{200}{\mega\hertz}. To filter out noise entering the sample through the output-line in the frequency range between \SI{4}{\giga\hertz} and \SI{8}{\giga\hertz}, we use microwave circulators at \SI{750}{\milli\kelvin} and at \SI{35}{\milli\kelvin}. The circulators have an average leakage of 0.02 photons at \SI{6}{\giga\hertz} due to their finite isolation. At the $4\,\mathrm{K}$ stage, we use a cryogenic high electron mobility transistor (HEMT) amplifier and further amplify the signal by a room temperature amplifier as shown in Fig.\,\ref{fig:setup}\,(b). We implement a time-resolved, phase-sensitive measurement of the in-phase and quadrature components $I(t)$ and $Q(t)$ of the readout signal by heterodyne downconversion to an intermediate frequency $\omega_{\mathrm{if}}/2\pi\,{=}\,\SI{62.5}{\mega\hertz}$ and subsequent amplification. We digitize the signals using two analog-to-digital converters (ADC) with a sampling rate of \SI{250}{\mega\hertz} and perform digital homodyning. In addition, we can read out the resonator via a vector network analyzer (VNA) for spectroscopic analysis of the sample. To generate pulsed sequences in the GHz regime, we mix a continuous microwave signal with a rectangular pulse generated by an arbitrary function generator (AFG). In addition to true thermal noise radiated from the attenuators, we can add noise generated by the AFG. By mixing this noise with a microwave signal, we upconvert the noise carrier frequency to the desired noise frequency $\omega_{\mathrm{n}}$. The AFG generates noise with a \SI{500}{\mega\hertz} bandwidth, which has a quasi-Gaussian amplitude distribution. This noise has a constant variance of \SI{1}{\volt} into \SI{50}{\ohm}, which we attenuate in a linear way when controlling the photon number. We additionally filter this noise before the upconversion to the carrier frequency $\omega_{\mathrm{n}}$ by two \SI{100}{\mega\hertz} low-pass filters. That way, the noise has a bandwidth of \SI{200}{\mega\hertz} and an on/off ratio of \SI{35}{\decibel} as shown in Fig.\,\ref{fig:scheme}\,(b).\\
\begin{figure}[t]
\includegraphics{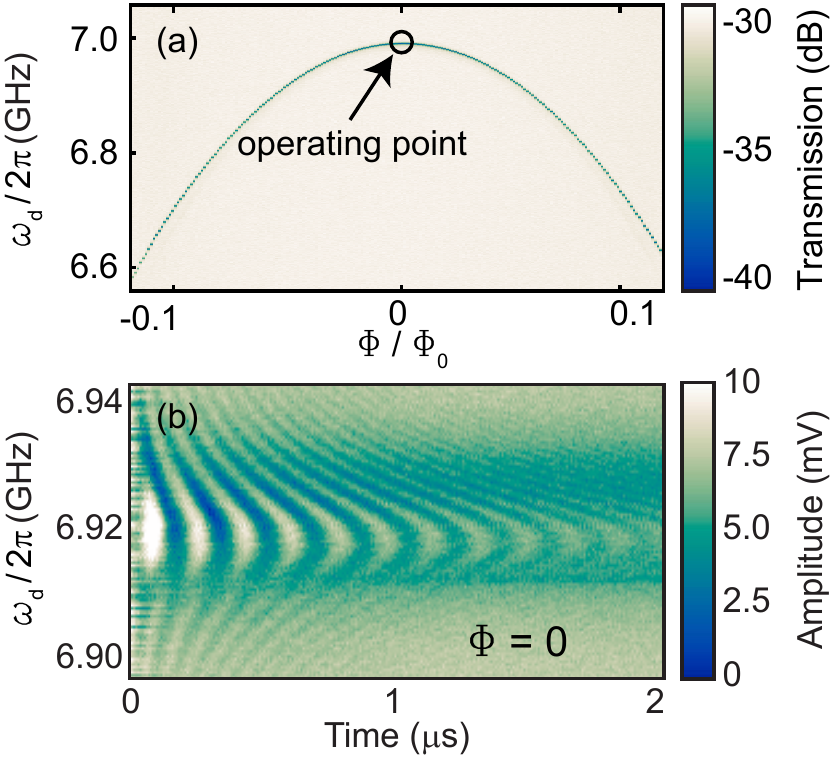}
\caption{\label{fig:wq} (a) Color encoded transmission magnitude plotted versus magnetic flux and drive frequency $\omega_{\mathrm{d}}$ measured in a continuous wave two-tone experiment. All measurements are performed at the flux sweet spot of the transmon qubit. (b) Driven Rabi oscillations encoded in the frequency dependent transmission amplitude plotted versus evolution time and drive frequency $\omega_{\mathrm{d}}$.}
\end{figure}
From a two-tone experiment, we extract the qubit transition frequency~\cite{Koch_2007} $\omega_{\mathrm{q}}\,{=}\,\omega_{\mathrm{q},0}\sqrt{\left|\cos\left(\pi\Phi/\Phi_{0}\right)\right|}$ by fitting a Lorentzian function to the dip in the transmission spectrum shown in Fig.\,\ref{fig:wq}\,(a). Here, $\Phi$ is the magnetic flux in the SQUID loop, which is generated by a superconducting coil outside the sample holder, and $\Phi_0$ is the flux quantum. All experiments are carried out in the dispersive regime by keeping the average resonator population $n_{\mathrm{r}}$ below the critical photon number~\cite{Gambetta_2006} $n_{\mathrm{crit}}\,{=}\,\delta^{2}/4g^{2}\,{\simeq}\,40$. In this limit, the system Hamiltonian $\mathcal{H}_{\mathrm{tot}}\,{=}\,\mathcal{H}_{\mathrm{r}}\,{+}\,\mathcal{H}_{\mathrm{q}}\,{+}\,\mathcal{H}_{\mathrm{int}}\,{+}\,\mathcal{H}_{\mathrm{d}}$ comprises the bare resonator Hamiltonian $\mathcal{H}_{\mathrm{r}}\,{=}\,\hbar\omega_{\mathrm{r}}n_{\mathrm{r}}$, the qubit Hamiltonian $\mathcal{H}_{\mathrm{q}}\,{=}\,\hbar\omega_{\mathrm{q}}\hat{\sigma}_{\mathrm{z}}/2$, the interaction Hamiltonian $\mathcal{H}_{\mathrm{int}}\,{=}\,\hbar\chi\left[n_{\mathrm{r}}\,{+}\,1/2\right]\hat{\sigma}_{\mathrm{z}}$, and a driving term $\mathcal{H}_{\mathrm{d}}\,{=}\,\hbar\Omega_{\mathrm{d}}\cos(\omega_{\mathrm{d}}t)\hat{\sigma}_{\mathrm{x}}$. Here, $\hat{\sigma}_{i}$ are the Pauli operators and $\Omega_{\mathrm{d}}$ defines the amplitude of the excitation drive with frequency $\omega_{\mathrm{d}}$. We use the qubit state dependent ac-Stark shift defined by $\mathcal{H}_{\mathrm{int}}$ for readout~\cite{Schuster_2005}. In Fig.\,\ref{fig:wq}\,(b), we show driven Rabi oscillations continuously recorded during a weak measurement using $n_{\mathrm{RO}}\,{\simeq}\,0.1$ readout photons.

\section{Coupling of thermal fields into a resonator}
\label{sec:broadband}

In this section, we derive how thermally induced voltage fluctuations on the feedlines influence the resonator population. Since we generate the thermal states outside the resonator, the mean photon population $n_{\mathrm{r}}$ inside the resonator can be calculated as a cavity field which is coupled to several bosonic baths each described by a Hamiltonian $\mathcal{H}_{\mathrm{bath}}\,{=}\,\sum^{\phantom{\dagger}}_{k}\hbar\omega^{\phantom{\dagger}}_{k}\hat{b}^{\dagger}_{k,j}\hat{b}^{\phantom{\dagger}}_{k,j}$. Here, the respective field operators $\hat{b}^{\dagger}_{k},\,(\hat{b}_{k})$ create (annihilate) the individual field modes with frequencies $\omega_{k}$. In our setup the  three bosonic reservoirs ($j\,{=}\,$i,x,a) couple to the resonator modes described by the operators $\hat{a},\,\hat{a}^{\dagger}$ via the interaction Hamiltonian $\mathcal{H}_{\mathrm{int}}\,{=}{-}\imath\hbar\sum_{k}[\kappa_{k,j}\hat{a}^{\dagger}\hat{b}_{k,j}\,{-}\,\kappa_{k,j}\hat{b}^{\dagger}_{k,j}\hat{a}]$. For convenience, we split up the transmission line modes into a classical part $\bar{b}_{k}$ originating from a coherent drive, and into a quantum part $\hat{\xi}_{k}$, such that $\hat{b}_{k}(t)\,{=}\,e^{-\imath\omega_{k}t}\bar{b}_{k}\,{+}\,\hat{\xi}_{k}(t)$~\cite{Devoret_2014}. The quantum part describes voltage fluctuations emitted from the heatable attenuators and will be the focus of the following discussion. The attenuators emit a voltage $V(t)\,{=}\,V_{\mathrm{vac}}[\hat{\xi}_{k}(t)\,{+}\,\hat{\xi}^{\dagger}_{k}(t)]$, which is fluctuating in time and has a Gaussian amplitude distribution. Here, $V_{\mathrm{vac}}$ is the vacuum amplitude of the corresponding mode. For a finite temperature, the correlation function for the voltage fluctuations is defined by the Hurwitz function~\cite{Kano_1962,Mehta_1963,Klein_2008}. The power spectral density of thermal fields $\mathcal{S}(\omega)$ in Eq.\,(\ref{eqn:SV2}) can then be obtained by a Fourier transform~\cite{Martinis_2003,Schoelkopf_1997} of the correlation function. We now discuss how thermal fields described by $\mathcal{S}(\omega)$ enter the resonator. It can be shown that the power spectrum inside the resonator is the product of the resonator modes and the modes entering from outside~\cite{Ujihara_1978}. Consequently, the relation between the field operator $\hat{a}$ inside the resonator and the input field reads~\cite{Walls_2008}
\begin{equation}
\hat{a}(\omega_{\mathrm{r}})\,{=}\,\sum_{j=\mathrm{i,x,ant}}\sum_{k}\frac{\sqrt{\kappa_{k,j}}\hat{b}^{\phantom{\dagger}}_{k,j}(\omega_{k})}{\kappa_{\mathrm{tot}}/2{-}\imath[\omega_{k}{-}\omega_{\mathrm{r}}]}\,.
\end{equation}
Due to the high density of modes, we take the continuum limit $(\sum_{k}\,{\mapsto}\,\int\mathrm{d}\omega_{k})$ and obtain the expression $n_{\mathrm{r}}(\omega,T)\,{\approx}\,\mathcal{F}_{\mathrm{L}}(\omega)\sum_{j}\kappa_{j}n_{j}(\omega,T)$ presented in Sec.\,\ref{sec:exp}. For a large qubit-resonator detuning $\delta\,{\gg}\,g,\kappa_{\mathrm{tot}}$, we can neglect the influence of qubit excitations entering the resonator. In this case, the Markovian master equation
\begin{align}
 \frac{\partial\hat{\rho}_{\mathrm{r}}}{\partial t} = &-\imath[\omega_{\mathrm{r}}\hat{a}^{\dagger}\hat{a},\hat{\rho}_{\mathrm{r}}]+\left\{n_{\mathrm{i}}\kappa_{\mathrm{i}}+n_{\mathrm{x}}\kappa_{\mathrm{x}}+n_{\mathrm{a}}\kappa_{\mathrm{a}}\right\}\mathcal{D}(\hat{a}^{\dagger})\hat{\rho}_{\mathrm{r}}\notag\\
&+\left\{(n_{\mathrm{i}}+1)\kappa_{\mathrm{i}}+(n_{\mathrm{x}}+1)\kappa_{\mathrm{x}}+(n_{\mathrm{a}}+1)\kappa_{\mathrm{a}}\right\}\mathcal{D}(\hat{a})\hat{\rho}_{\mathrm{r}}\,,
\label{eqn:markovres}
\end{align}
describes the resonator, where $\hat{\rho}_{\mathrm{r}}$ is the density matrix of the undisturbed resonator and~$\mathcal{D}(\hat{L})$ is the Lindblad operator. In the steady state limit, Eq.\,(\ref{eqn:markovres}) becomes
\begin{equation}
n_{\mathrm{r}} = \frac{n_{\mathrm{i}}\kappa_{\mathrm{i}}+n_{\mathrm{x}}\kappa_{\mathrm{x}}+n_{\mathrm{a}}\kappa_{\mathrm{a}}}{\kappa_{\mathrm{i}}+\kappa_{\mathrm{x}}+\kappa_{\mathrm{a}}}\,.
\label{eqn:ada}
\end{equation}
This equation shows that we can precisely control the power inside the resonator using thermal photons emitted from the heatable attenuators. Concerning power, the nature of the photons inside the resonator (e.g., thermal or coherent) makes no difference.

\section{First- and second-order coupling between thermal fields and the qubit}
\label{sec:2ndordercoupling}

In this section, we derive the first-order and second-order dephasing rates for the transmon qubit due to thermal fields on on-chip control lines.

\begin{figure}[t]
\includegraphics{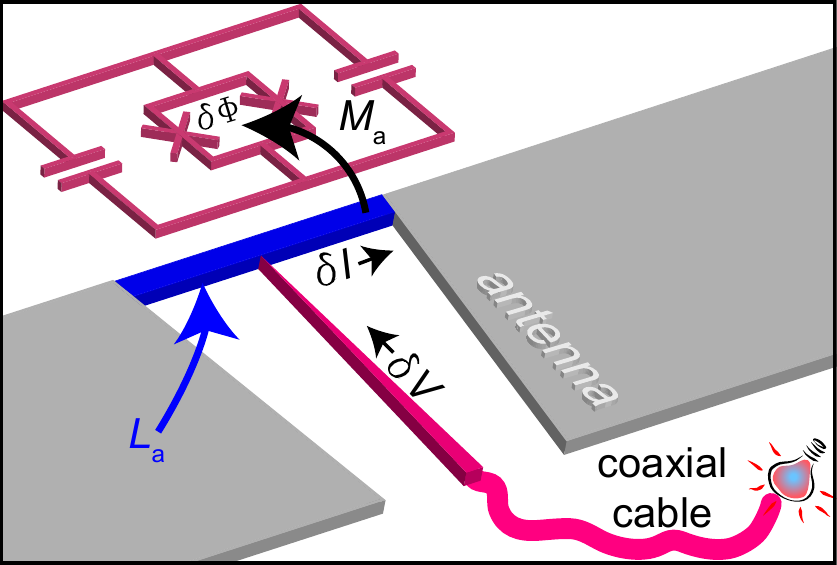}
\caption{\label{fig:coupling}Sketch of the short-circuited antenna line, which converts voltage fluctuations $\delta V$ into flux fluctuations $\delta\Phi$ inside the SQUID loop of the transmon. The region forming the short-circuit has an inductance $L_{\mathrm{a}}$ and a mutual inductance $M_{\mathrm{a}}$ to the SQUID loop. The propagating thermal fields are guided through a coaxial cable to the sample and are thus only located on the antenna structure.}
\end{figure}

\paragraph*{First-order coupling} First-order coupling between flux fluctuations $\delta\lambda\,{\equiv}\,\delta\Phi/\Phi_{0}$ and the qubit follow the Hamiltonian $\mathcal{H}_{\mathrm{sys}}\,{=}\,[\hbar/2][\omega_{\mathrm{q}}\hat{\sigma}_{\mathrm{z}}\,{+}\,\delta\omega_{\mathrm{q}}\hat{\sigma}_{\mathrm{z}}]$. Here, $\delta\omega_{\mathrm{q}}\,{=}\,\delta\lambda\mathcal{D}^{(1)}_{\lambda,\mathrm{z}}$ describes fluctuations of the qubit transition frequency leading to dephasing characterized by the first-order transfer function
\begin{equation}
 \mathcal{D}^{(1)}_{\lambda,\mathrm{z}}(\lambda^{\star}) \equiv \frac{1}{\hbar}\left.\frac{\partial\mathcal{H}_{\mathrm{q}}(\lambda)}{\partial\lambda}\right|_{\lambda^{\star}} =  -\left. \frac{\pi\omega_{\mathrm{q},0}}{2}\frac{\sin\left(\pi\lambda\right)}{\sqrt{\cos\left(\pi\lambda\right)}}\right|_{\lambda^{\star}}.
\label{eqn:t1st}
\end{equation}
Equation (\ref{eqn:t1st}) is defined at flux operating points $\lambda^{\star}\,{\in}\,[{-}1/2,\,1/2]$ using the transmon qubit Hamiltonian $\mathcal{H}_{\mathrm{q}}\,{=}\,\hbar\omega_{\mathrm{q},0}\sqrt{|\cos(\pi\lambda)|}$. To analyze the fluctuations $\delta\omega_{\mathrm{q}}$, we first derive how voltage fluctuations $\delta V$ on the antenna line are converted into flux fluctuations $\delta\lambda$ in the SQUID loop. Because the antenna is short-circuited near the qubit by a finite inductance $L_{\mathrm{a}}$ as depicted in Fig.\,\ref{fig:coupling}, we describe it as a first-order low-pass $LR$ filter, i.e., $Z_{\mathrm{a}}(\omega)\,{\approx}\,\omega^{2}L_{\mathrm{a}}^{2}/Z_{0}$ (similar to the way presented in Ref.\,\onlinecite{vanderWal_2003}). The finite inductance $L_{\mathrm{a}}$ of the short-circuit converts voltage fluctuations $\delta V$ emitted from the attenuator into current fluctuations $\delta I\,{=}\,\delta V(\imath\omega L_{\mathrm{a}})^{-1}$. Via the mutual inductance $M_{\mathrm{a}}$ between antenna and SQUID loop, these current fluctuations cause flux fluctuations $\delta\lambda\,{=}\,M_{\mathrm{a}}\delta I/\Phi_{0}$. Using the transfer function in Eq.\,(\ref{eqn:t1st}), we calculate the resulting change in transition frequency to
\begin{align}
  \delta\omega_{\mathrm{q}}(\lambda^{\star}) &=\mathcal{D}^{(1)}_{\lambda,\mathrm{z}}(\lambda^{\star})\times\delta\lambda\notag\\
  &= \mathcal{D}^{(1)}_{\lambda,\mathrm{z}}(\lambda^{\star})\frac{M_{\mathrm{a}}}{\Phi_{0}}\frac{\delta V}{\imath\omega L_{\mathrm{a}}}\,.
\label{eqn:deltaomegaq}
\end{align}

This equation shows that thermally induced voltage fluctuations indeed lead to fluctuations $\delta\omega_{\mathrm{q}}$ of the qubit transition frequency. To calculate the resulting dephasing rate we use the spectral function
\begin{align}
  \langle\delta\omega_{\mathrm{q}}(t)\delta\omega_{\mathrm{q}}(0)\rangle_{\omega} &= \left[\frac{\mathcal{D}^{(1)}_{\lambda,\mathrm{z}}(\lambda^{\star})M_{\mathrm{a}}}{\omega L_{\mathrm{a}}\Phi_{0}}\right]^{2}\langle\delta V(t)\delta V(0)\rangle_{\omega}\notag\\
 &= \left[\frac{\mathcal{D}^{(1)}_{\lambda,\mathrm{z}}(\lambda^{\star})M_{\mathrm{a}}}{\omega L_{\mathrm{a}}\Phi_{0}}\right]^{2}\mathrm{Re}\{Z_{\mathrm{a}}(\omega)\}\mathcal{S}(\omega)\notag\\
  &= \left[\mathcal{D}^{(1)}_{\lambda,\mathrm{z}}(\lambda^{\star})\frac{M_{\mathrm{a}}}{\Phi_{0}}\right]^{2}\frac{\mathcal{S}(\omega)}{Z_{0}}\,
\label{egn:domega}
\end{align}
of these fluctuations. Here, we use the fluctuation dissipation theorem $\langle\delta V(t)\delta V(0)\rangle_{\omega}\,{=}\,\mathrm{Re}\{Z_{\mathrm{a}}(\omega)\}\mathcal{S}(\omega)$~\cite{Nyquist_1928,Callen_1951} to relate the voltage fluctuations to their power spectral density. The expression $\langle\delta V(t)\delta V(0)\rangle_{\omega}$ displays the spectral weight of the fluctuations with units \SI{}{\square\volt\per\hertz}. The fluctuations of the qubit transition frequency defined in Eq.\,(\ref{egn:domega}) lead to random fluctuations $\delta\varphi(t)\,{=}\,\int_{0}^{t}\mathrm{d} t^{\prime}\delta\omega_{\mathrm{q}}(t^{\prime})$ of the qubit phase relative to its mean phase $\bar{\varphi}$ in a rotating frame. Assuming thermal states to be Gaussian~\cite{Shnirman_2002,Makhlin_2004} and $1/f$ contributions to be negligible, the phase fluctuations in turn lead to a single exponential decay function~\cite{Gambetta_2006}
\begin{align}
 \langle\hat{\sigma}_{-}(t)&\hat{\sigma}_{+}(0)\rangle \approx \exp\left[{-}\gamma_{2}t{-}\langle\delta\varphi^{2}\rangle/2\right]\notag\\ &= \exp\left[{-}\gamma_{2}t-\frac{1}{2}\int\int_{0}^{t}\mathrm{d} t_{1}\mathrm{d} t_{2}\langle\delta\omega_{\mathrm{q}}(t_{1})\delta\omega_{\mathrm{q}}(t_{2})\rangle\right]\notag\\
 &= \exp\left[{-}\gamma_{2}t-\frac{1}{2}\left[\mathcal{D}^{(1)}_{\lambda,\mathrm{z}}(\lambda^{\star})\frac{M_{\mathrm{a}}}{\Phi_{0}}\right]^{2}\frac{\mathcal{S}(\omega)}{Z_{0}} t\right]\,.
\label{eqn:decoh1}                                                                                                                                                                                                                                                                                                                                                                                                                                                                                                                                                                                                                                                         \end{align}
Here, we make use of the fact that the correlation function $\langle\delta\omega_{\mathrm{q}}(t_{1})\delta\omega_{\mathrm{q}}(t_{2})\rangle$ is a $\delta$-function at low frequencies. The additional decoherence due to thermal states on the antenna line is therefore given as
\begin{equation}
 \gamma_{\varphi}^{(1)} = \left[\mathcal{D}^{(1)}_{\lambda,\mathrm{z}}(\lambda^{\star})\frac{M_{\mathrm{a}}}{\Phi_{0}}\right]^{2}\frac{\mathcal{S}(\omega)}{2Z_{0}} \approx \left[\mathcal{D}^{(1)}_{\lambda,\mathrm{z}}(\lambda^{\star})\frac{M_{\mathrm{a}}}{\Phi_{0}}\right]^{2}\frac{k_{\mathrm{B}}T_{\mathrm{a}}}{Z_{0}}\,.
\label{eqn:gammaphi1}
\end{equation}
The approximation on the right-hand side of Eq.\,(\ref{eqn:gammaphi1}) reflects the low-frequency limit by setting $\mathcal{S}(\omega{\mapsto}0)\,{=}\,k_{\mathrm{B}}T_{\mathrm{a}}$, which shows that dephasing due to propagating thermal fields is expected to increase linearly with the temperature $T_{\mathrm{a}}$ of the black-body radiator.\\
The above derivation of the qubit dephasing rate is equivalent to the result derived from the spin-boson model, where the dephasing is defined as~\cite{vanderWal_2003} $\gamma_{\varphi}^{(1)}\,{=}\,2\pi\alpha\mathcal{S}(\omega{\mapsto}0)/\hbar$. Here, the dimensionless dissipation parameter $\alpha$ is defined as~\cite{Shnirman_2002}
\begin{equation}
 \alpha \equiv \frac{R_{\mathrm{q}}}{Z_{0}}\left[\hbar\mathcal{D}^{(1)}_{\lambda,\mathrm{z}}(\lambda^{\star})\frac{\partial\lambda}{\partial\Phi}\frac{M_{\mathrm{a}}}{\Phi_{0}}\right]^{2} = \frac{\hbar}{2\pi Z_{0}}\left[\mathcal{D}^{(1)}_{\lambda,\mathrm{z}}(\lambda^{\star})\frac{M_{\mathrm{a}}}{\Phi_{0}}\right]^{2}
\end{equation}
and $R_{\mathrm{q}}\,{\equiv}\,h/4e^{2}$ is the resistance quantum for Cooper pairs. We note that the spin-boson model can be applied to calculate the dephasing rate because the antenna creates an ohmic environment if modeled as an $LR$-filter and because thermal noise has no $1/f$ contribution.

\paragraph*{Second-order coupling} When the transmon qubit is operated at the flux sweet spot, dephasing can be dominated from intensity fluctuations coupling in second-order to the qubit. These second-order fluctuations
\begin{equation}
 \delta\lambda^{2} \equiv \frac{\delta\Phi^{2}}{\Phi_{0}^{2}} = \frac{M_{\mathrm{a}}^{(2)}\delta I^{2}}{\hbar\omega_{\mathrm{q},0}} =  \frac{M_{\mathrm{a}}^{2}}{L_{\ell}}\frac{\delta I^{2}}{\hbar\omega_{\mathrm{q},0}}
\end{equation}
are normalized to the relevant energy scale $\hbar\omega_{\mathrm{q},0}$ of the qubit and scale with the second-order mutual inductance $M_{\mathrm{a}}^{(2)}\,{=}\,M_{\mathrm{a}}^{2}/L_{\ell}$ due to the inductive energy $E_{\ell}\,{=}\,\Phi_{0}^{2}/2L_{\ell}$ of the SQUID loop with inductance $L_{\ell}$. The intensity fluctuations induce fluctuations $\delta\omega_{\mathrm{q}}\,{=}\,\mathcal{D}^{(2)}_{\lambda,\mathrm{z}}\delta\lambda^{2}/2$ of the qubit transition frequency leading to dephasing based on the system Hamiltonian $\mathcal{H}_{\mathrm{sys}}$ defined in the previous paragraph. The frequency fluctuations are characterized by the second-order derivative of the qubit transition frequency
\begin{align}
 &\mathcal{D}^{(2)}_{\lambda,\mathrm{z}}(\lambda^{\star}) \equiv \frac{1}{\hbar}\left.\frac{\partial^{2}\mathcal{H}_{\mathrm{q}}(\lambda)}{\partial\lambda^{2}}\right|_{\lambda^{\star}}\notag\\
& =   -\frac{\pi^{2}\omega_{\mathrm{q},0}}{2}\sqrt{\cos(\pi\lambda)} - \left. \frac{\pi^{2}\omega_{\mathrm{q},0}}{4}\frac{\sin^{2}(\pi\lambda)}{\cos^{3/2}(\pi\lambda)}\right|_{\lambda^{\star}}\,.
\label{eqn:t2nd}
\end{align}
Using this transfer function as well as $\delta V^{2}\,{=}\,(\omega L_{\mathrm{a}})^{2}\delta I^{2}$, we can characterize fluctuations in the qubit transition frequency similar to Eq.\,(\ref{egn:domega}) as
\begin{align}
  \langle\delta\omega_{\mathrm{q}}(t)&\delta\omega_{\mathrm{q}}(0)\rangle_\omega = \left[\frac{\mathcal{D}^{(2)}_{\lambda,\mathrm{z}}(\lambda^{\star})}{2}\right]^{2} \langle\delta\lambda^{2}(t)\delta\lambda^{2}(0)\rangle_\omega\,\notag\\
 &= \left[\frac{\mathcal{D}^{(2)}_{\lambda,\mathrm{z}}(\lambda^{\star})}{2}\frac{M_{\mathrm{a}}^{2}}{L_{\ell}}\frac{1}{\omega^{2}L_{\mathrm{a}}^{2}}\right]^{2} \frac{\langle\delta V^{2}(t)\delta V^{2}(0)\rangle_\omega}{(\hbar\omega_{\mathrm{q},0})^{2}}\,\notag\\
   &= \left[\frac{\mathcal{D}^{(2)}_{\lambda,\mathrm{z}}(\lambda^{\star})}{2}\frac{M_{\mathrm{a}}^{2}}{L_{\ell}}\frac{1}{\omega^{2}L_{\mathrm{a}}^{2}}\right]^{2}\mathrm{Re}\{Z_{\mathrm{a}}(\omega)\}^{2}\frac{\mathcal{S}^{(2)}(\omega)}{(\hbar\omega_{\mathrm{q},0})^{2}}\,\notag\\
&=\left[\frac{\mathcal{D}^{(2)}_{\lambda,\mathrm{z}}(\lambda^{\star})}{2}\frac{M_{\mathrm{a}}^{2}}{L_{\ell}}\right]^{2}\frac{\mathcal{S}^{(2)}(\omega)}{(\hbar\omega_{\mathrm{q},0})^{2}Z_{0}^{2}}\,.
\label{eqn:domega2}
\end{align}
where $\mathcal{S}^{(2)}(\omega)$ is the second-order spectral density defined in Eq.\,(\ref{eqn:Sth2}). To derive the dephasing rate using Eq.\,(\ref{eqn:domega2}), we have to consider the statistical properties of $\mathcal{S}^{(2)}(\omega)$. Because the intensity fluctuations are not Gaussian distributed~\cite{Shnirman_2002}, applying an approach similar to the one in Eq.\,(\ref{eqn:decoh1}) to calculate the dephasing rate from second-order thermal fields is not valid in general. Performing only first-order pertubative analysis, we can, however, assume that the second-order fluctuations are Gaussian distributed with a width that is defined by $\mathcal{S}^{(2)}(\omega)$. Then, we find the additional dephasing rate due to second-order noise
\begin{align}
 \gamma_{\varphi\mathrm{,a}}^{(2)} &= \left[\frac{\mathcal{D}^{(2)}_{\lambda,\mathrm{z}}(\lambda^{\star})}{2}\frac{M_{\mathrm{a}}^{2}}{L_{\ell}}\right]^{2}\frac{\mathcal{S}^{(2)}(\omega)}{(\hbar\omega_{\mathrm{q},0})^{2}Z_{0}^{2}} \notag\\
&\approx \underbrace{\left[\frac{\mathcal{D}^{(2)}_{\lambda,\mathrm{z}}(\lambda^{\star})}{2\sqrt{3}}\frac{M_{\mathrm{a}}^{2}}{L_{\ell}Z_{0}}\right]^{2}}_{\alpha^{(2)}} \underbrace{\left[\frac{k_{\mathrm{B}}T_{\mathrm{a}}}{\hbar\omega_{\mathrm{q},0}}\right]^{2}}_{r}\frac{2\pi k_{\mathrm{B}}T_{\mathrm{a}}}{\hbar}\,,
 \label{eqn:gammaphi2}
\end{align}
taking the low-frequency limit $\mathcal{S}^{(2)}(\omega{\mapsto}0)\,{\approx}\,2\pi k_{\mathrm{B}}^{3}T_{\mathrm{a}}^{3}/3\hbar$ obtained from Eq.\,(\ref{eqn:Sth2}). Just as the first-order dephasing rate, the second-order dephasing rate is equivalent to an approach based on the spin-boson model when using the second-order dissipation factor $\alpha^{(2)}$. When comparing Eq.\,(\ref{eqn:gammaphi2}) to Eq.\,(\ref{eqn:gammaphi1}), we see that second-order thermal noise is suppressed by the factor $r$ if the thermal energy is lower than the qubit energy. Hence, we have to use temperatures $T_{\mathrm{a}}\,{>}\,\hbar\omega_{\mathrm{q},0}/k_{\mathrm{B}}$ or work at the flux sweet spot to observe the $T^{3}$ law.

\bibliography{Goetz_Bibliography}

\end{document}